\documentclass[
 amsmath,amssymb,
 aps,
prb,
floatfix
]{revtex4-2}

\usepackage{setspace} 
\usepackage{graphicx}
\usepackage{upgreek}
\usepackage{amsmath}
\usepackage{dcolumn}
\usepackage{float}
\usepackage[table]{xcolor}
\usepackage{bm}
\usepackage{comment}

\begin{document}
\title{Scanning SQUID study of ferromagnetism and superconductivity in infinite-layer nickelates}

\author{Ruby A. Shi$^{1,2,3}$}
\author{Bai Yang Wang$^{1,2}$}
\author{Yusuke Iguchi$^{1,3}$}
\author{Motoki Osada$^{1,4}$}
\author{Kyuho Lee$^{1,2}$} 
\author{Berit H. Goodge$^{6,7}$}
\author{Lena F. Kourkoutis$^{6,7}$}
\author{Harold Y. Hwang$^{1,3,5}$}
\author{Kathryn A. Moler$^{1,2,3,5}$}

\affiliation{$^1$Stanford Institute for Materials and Energy Sciences, SLAC National Accelerator Laboratory, 2575 Sand Hill Road, Menlo Park, California 94025-7015, USA}
\affiliation{$^2$Department of Physics, Stanford University, California 94305-4045, USA}
\affiliation{
$^3$Geballe Laboratory for Advanced Materials, Stanford University, Stanford, California 94305-4045, USA
}
\affiliation{$^4$Department of Material Science and Engineering, Stanford University, Stanford, California 94305-4045, USA}
\affiliation{$^5$Department of Applied Physics, Stanford University, California 94305-4045, USA}
\affiliation{$^6$School of Applied and Engineering Physics, Cornell University, Ithaca, New York 14853-3501, USA}
\affiliation{$^7$Kavli Institute at Cornell for Nanoscale Science, Cornell University, Ithaca, New York 14853-3501, USA}


\begin{abstract}
Infinite-layer nickelates $R_{1-x}$Sr$_{x}$NiO$_{2}$ ($R$ = La, Pr, Nd) are a class of superconductors with structural similarities to cuprates. Although long-range antiferromagnetic order has not been observed for these materials, magnetic effects such as antiferromagnetic spin fluctuations and spin-glass behavior have been reported. Different experiments have drawn different conclusions about whether the pairing symmetry is $s$- or $d$ wave. In this paper, we applied a scanning superconducting quantum interference device (SQUID) to probe the magnetic behavior of film samples of three infinite-layer nickelates (La$_{0.85}$Sr$_{0.15}$NiO$_2$, Pr$_{0.8}$Sr$_{0.2}$NiO$_2$, and Nd$_{0.775}$Sr$_{0.225}$NiO$_2$) grown on SrTiO$_3$ (STO), each with a nominal thickness of 20 unit cells. In all three films, we observed a ferromagnetic background. We also measured the magnetic susceptibility above the superconducting critical temperature in Pr$_{0.8}$Sr$_{0.2}$NiO$_2$ and La$_{0.85}$Sr$_{0.15}$NiO$_2$ and identified a non-Curie-Weiss dynamic susceptibility. Both magnetic features are likely due to NiO$_x$ nanoparticles. Additionally, we investigated superconductivity in Pr$_{0.8}$Sr$_{0.2}$NiO$_2$ and Nd$_{0.775}$Sr$_{0.225}$NiO$_2$, which exhibited inhomogeneous diamagnetic screening. The superfluid density inferred from the diamagnetic susceptibility in relatively homogeneous regions shows $T$-linear behavior in both samples. Finally, we observed superconducting vortices in Nd$_{0.775}$Sr$_{0.225}$NiO$_2$. We determined a Pearl length of 330 $\upmu$m for Nd$_{0.775}$Sr$_{0.225}$NiO$_2$ at 300 mK both from the strength of the diamagnetism and from the size and shape of the vortices. These results highlight the importance of considering NiO$_x$ particles when interpreting experimental results for these films.  
\end{abstract}


\maketitle


Since their discovery \cite{discovery}, superconducting infinite-layer nickelates have attracted tremendous attention \cite{Botana_2021, ChineseRev, JapaneseRev, FrontierYang, FrontierMitchell}. These nickelates are structural analogs of cuprates; however, they exhibit electronic and magnetic differences \cite{FrontierBotana}. Transport and spectroscopy investigations have revealed their multiband electronic nature \cite{SrDome, PrDome, Goodge, XRay, ARPES, FrontierHeld, FrontierHirayama}. Theoretical calculations of the antiferromagnetic exchange coupling $J_1$ range from very weak, at 10 meV \cite{AFM10meV,2020AFMtraces}, up to 100 meV \cite{2020AFMUniversal, 2021AFM,FrontierChen,AFM82meV,AFM100meV,AFM77meV,AFMBotana100}. Nuclear magnetic resonance (NMR) on powdered Nd$_{0.85}$Sr$_{0.15}$NiO$_2$ \cite{NMR}  has indicated short-range antiferromagnetic fluctuations, obtaining $J_1$ on the order of 10 meV. Similarly, Raman spectroscopy on bulk NdNiO$_2$ \cite{raman}, and resonance inelastic x-ray scattering (RIXS) on thin-film NdNiO$_2$ \cite{RIXS}  have demonstrated $J_1$ values of 25 and 63.6 meV. However, there has been no evidence reported of long-range antiferromagnetic order. Muon spin rotation/relaxation ($\upmu$SR) has demonstrated multiple intrinsic magnetic orders in different temperature ranges \cite{muSR}. Some theories predict $d$-wave pairing symmetry of the superconducting order parameter, like cuprates \cite{DWavePredict, DWaveJpn, DWavePredict2, PossibleD}, yet other calculations have suggested that the doping level and surface termination might result in $s$-wave or ($s$+$d$)-wave pairing\cite{SWaveDoping, SWaveSurface}. Tunnel-diode-oscillator and single-particle tunneling spectroscopy studies on various samples have shown $s$-wave pairing in Nd$_{0.8}$Sr$_{0.2}$NiO$_2$, ($s$+$d$)-wave pairing in La$_{0.8}$Ca$_{0.2}$NiO$_2$ \cite{sing_SD}, and a spatially inhomogeneous mixture of $s$-wave and $d$-wave pairing in Nd$_{0.8}$Sr$_{0.2}$NiO$_2$ \cite{2020Tunnel}. In contrast, two-coil mutual-inductance measurements of the superfluid density $n_s$ have demonstrated $T$-linear behavior at low temperatures indicating the presence of nodes consistent with $d$-wave pairing symmetry in La$_{0.8}$Sr$_{0.2}$NiO$_{2}$ and Pr$_{0.8}$Sr$_{0.2}$NiO$_{2}$ \cite{Hwang_SD}. 
These seemingly inconsistent results raise the question of whether the magnetism and $n_s$ are  homogeneous. 
\par
In this paper, we have applied a scanning superconducting quantum interference device (SQUID) to study the local magnetic response of three thin-film nickelates, $R_{1-x}$Sr$_{x}$NiO$_{2}$ ($R$ = La, Pr, Nd), grown by pulsed laser deposition on SrTiO$_3$ (STO). First, we discuss the main magnetic features that we observe: diamagnetism, as expected for superconductors, and an inhomogeneous ferromagnetic background. Second, we explain the basis for attributing this ferromagnetic background to NiO$_x$ particles. Third, we analyze the dynamic diamagnetic susceptibility to extract the Pearl length and its temperature dependence. Finally, we further investigate the static magnetic signal and show that, by subtracting the ferromagnetic background, we can observe superconducting vortices, which is a necessary condition for further phase-sensitive studies in tricrystals for determining the superconducting order parameter \cite{triCrystal}. 
\par
Our SQUIDs are capable of revealing magnetic impurities and diamagnetic screening inhomogeneities. The SQUID gradiometer used in this paper has a circular pickup loop with an inner diameter of 6 $\upmu$m and a concentric field coil that provides a local magnetic field \cite{John2016}. The pickup loop records the static magnetic field of the sample [direct current (DC) magnetometry in units of flux quanta $\Phi_0 = hc/2e$] and enables two-coil measurements with a current through the field coil [alternating current (AC) susceptometry in flux quanta per ampere $\Phi_0/A$]. Previous works have demonstrated that our SQUIDs can measure magnetic dipoles \cite{LAOSTO}, vortices \cite{PearlVortex}, flux quantization \cite{triCrystal}, and inhomogeneous diamagnetic screening \cite{Sam}, among other phenomena. By identifying local (tens of microns) regions with more homogeneous screening, we can determine the local $n_s$ in relatively homogeneous regions of the film.  
\par
We studied one control sample and three $\sim$6.7-nm-thick nickelate films (La$_{0.85}$Sr$_{0.15}$NiO$_2$, Pr$_{0.8}$Sr$_{0.2}$NiO$_2$, and Nd$_{0.775}$Sr$_{0.225}$NiO$_2$) on STO. All four samples, including the control sample, have a few-layer STO cap and have gone through topotactic reduction, as described in Ref. \cite{2020Synthesis}. Electronic transport characterizations on these samples confirmed zero-resistance superconducting critical temperatures of 5.1 and 7.8 K for Nd$_{0.775}$Sr$_{0.225}$NiO$_2$ and Pr$_{0.8}$Sr$_{0.2}$NiO$_2$, respectively. Here, La$_{0.85}$Sr$_{0.15}$NiO$_2$ has an onset resistance drop at $\sim$ 8 K, but does not show a zero-resistance transition until 2 K (see the Supplemental Material \cite{Supplemental}).   
\begin{figure}[htb]
\centering
  \includegraphics[scale = .9]{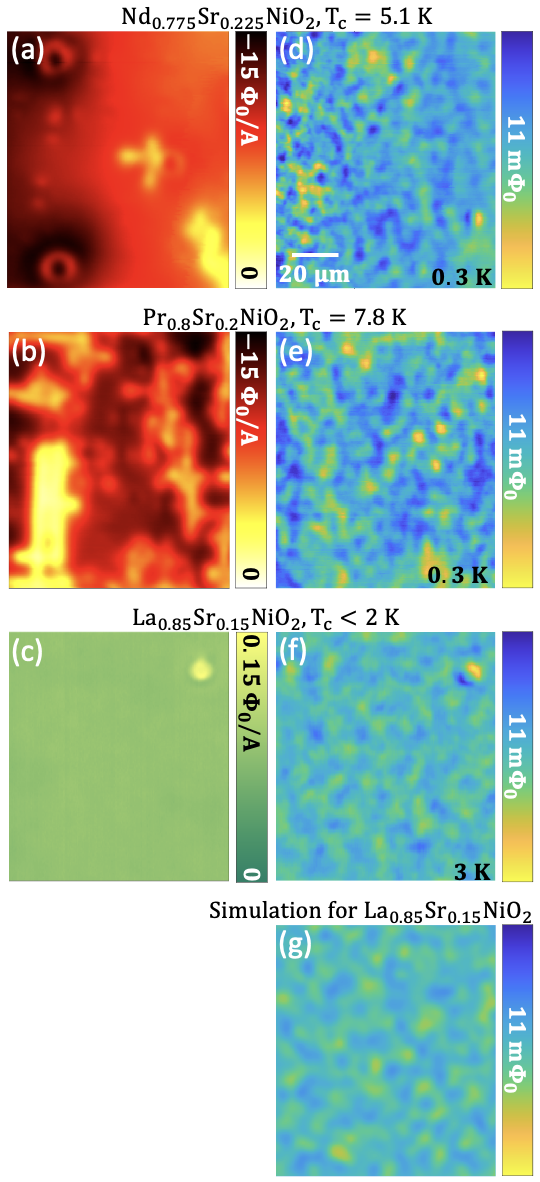}
  \caption{Diamagnetic screening in the presence of a ferromagnetic background. Susceptometry images of (a) Nd$_{0.775}$Sr$_{0.225}$NiO$_2$ ($T=0.3$ K), (b) Pr$_{0.8}$Sr$_{0.2}$NiO$_2$ ($T=0.3$ K), and (c) La$_{0.85}$Sr$_{0.15}$NiO$_2$ ($T=3$ K). The images of  Nd$_{0.775}$Sr$_{0.225}$NiO$_2$ and Pr$_{0.8}$Sr$_{0.2}$NiO$_2$, which are below their respective critical temperatures, show diamagnetic screening, while the image of La$_{0.85}$Sr$_{0.15}$NiO$_2$, which is above its critical temperature, shows paramagnetism. (d)--(f)  Magnetometry images taken simultaneously with the susceptometry images. A planar background fit is subtracted from each scan. The variation in the sharpness of the image from left to right in (a) is due to an imperfect scan plane. The estimated scan heights are 4.5 $\upmu$m (Nd$_{0.775}$Sr$_{0.225}$NiO$_2$), 3.1 $\upmu$m (Pr$_{0.8}$Sr$_{0.2}$NiO$_2$), and 3.1 $\upmu$m (La$_{0.85}$Sr$_{0.15}$NiO$_2$) with the bender voltage of 1~V(see the Supplemental Information \cite{Supplemental}). (g) Simulated magnetometry image of randomly positioned in-plane magnetic dipoles each carrying a magnetic moment of $m_0 = 240 000 \upmu_B$, with a density of 37.5 /$\upmu$m$^2$, at a scan height of 3.1 $\upmu$m. 
}
\label{main_coexist}
\end{figure}

Figure~\ref{main_coexist} presents images of AC magnetic susceptibility [Figs.~\ref{main_coexist}(a)--(c)] and DC magnetometry [Figs.~\ref{main_coexist}(d)--(f)] for the three nickelate films. 

In Nd$_{0.775}$Sr$_{0.225}$NiO$_2$ and Pr$_{0.8}$Sr$_{0.2}$NiO$_2$ at $T=0.3$ K, below the superconducting critical temperatures $T_c$, we observed diamagnetic screening, as expected [Figs.~\ref{main_coexist}(a) and \ref{main_coexist}(b)]. (In this paper, we use the convention that diamagnetic screening is reported in negative $\Phi_0/A$ and is plotted in a red--yellow colormap [Figs.~\ref{main_coexist}(a) and \ref{main_coexist}(b)], while paramagnetic screening  is reported in positive $\Phi_0/A$ and is plotted in a green--yellow colormap [Fig.~\ref{main_coexist}(c)].) The diamagnetic screening in Nd$_{0.775}$Sr$_{0.225}$NiO$_2$ and Pr$_{0.8}$Sr$_{0.2}$NiO$_2$ shows two types of inhomogeneities: cross-shaped features in both samples [Figs.~\ref{main_coexist}(a) and \ref{main_coexist}(b)] and donut-shaped features in otherwise relatively homogeneous regions of Nd$_{0.775}$Sr$_{0.225}$NiO$_2$ [Fig.~\ref{main_coexist}(a)]. The cross-shaped features are similar in size and shape to features that are visible in optical microscopy across all superconducting nickelates (see the Supplemental Information \cite{Supplemental}). The donut-shaped features in  Nd$_{0.775}$Sr$_{0.225}$NiO$_{2}$ indicate a point-like feature with greatly reduced diamagnetic response in the film, convolved with our sensor geometry \cite{defect}. 

In La$_{0.85}$Sr$_{0.15}$NiO$_2$ at $T=3$ K [Fig.~\ref{main_coexist}(c)], above $T_c$, we see a relatively uniform paramagnetic response. There is a small feature in the upper right corner of the image that shows a larger paramagnetic response. This feature may be due to a small contaminating particle on the surface. 

Surprisingly, we see an inhomogeneous ferromagnetic background in all three films [Figs.~\ref{main_coexist}(d)--~\ref{main_coexist}(f)]. The absence of similar inhomogeneous ferromagnetism in the control substrate, as shown in the Supplemental Material \cite{Supplemental}, indicates that the ferromagnetic background originates from the nickelate film. 
\par
To investigate the origin of the ferromagnetic background, we conducted atomic force microscopy and transmission electron microscopy. Atomic force microscopy demonstrated submicron-sized particles on the surfaces of each measured nickelate sample. Cross-sectional scanning transmission electron microscopy on Nd$_{0.775}$Sr$_{0.225}$NiO$_2$ identified these particles as NiO$_x$ nanoparticles at the interface between the nickelate film and the STO cap \cite{Supplemental}. Bulk NiO is an antiferromagnet, yet it is well established that NiO$_x$ nanoparticles can be single-domain ferromagnets with a magnetic moment of $m_0$ \cite{2004NiO,2005NiO,2006NiO,2009NiO, NiO2015} . 

Therefore, we use a random dipole model to model the shape and magnitude of the observed features. We assume 2.8 Bohr magneton ($\mu_B$) for each NiO$_x$ unit, based on three recent reports \cite{2006NiO, 2009NiO, NiO2015}, giving each  nanoparticle in the model a magnetic moment of $m_0$ = 240,000 $\upmu_B$ (see the Supplemental Information ~\cite{Supplemental}).  The model also assumes 37.5 nanoparticles/$\upmu$m$^2$, based on atomic force microscopy, with random locations. We model each NiO$_x$ particle as a single dipole of $m_0$ with a random in-plane orientation and a random $(x, y)$ location in the simulation area. Figure~\ref{main_coexist}(g) presents simulation results at the fitted scanning height, as discussed below, for the data shown for La$_{0.85}$Sr$_{0.15}$NiO$_2$ [Fig.~\ref{main_coexist}(f)]. The simulation qualitatively captures the magnitude and spatial size of the observed ferromagnetic features. 
\par
Based on our observations of NiO$_x$ nanoparticles and the agreement between the experimental ferromagnetic background data and a model based on NiO$_x$ nanoparticles, the ferromagnetic background may be due to extrinsic NiO$_x$ nanoparticles.  
\par
\begin{figure}[htb]
\centering
  \includegraphics[scale = .75]{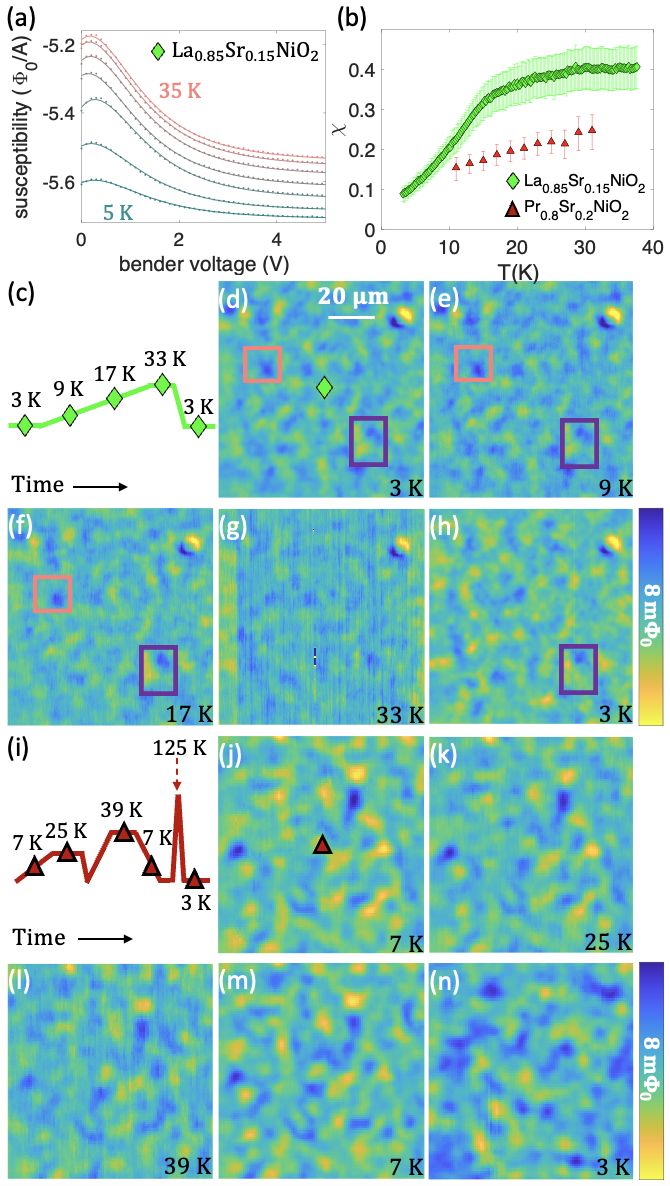}
  \caption{AC susceptibility and DC magnetometry consistent with superparamagnetism in extrinsic nanomagnetic particles. (a) Susceptometry vs. bender voltage at 5, 10, 15, 20, 25, 30, and 35 K, fitted to the expected functional for a thin-film paramagnet. The bender voltage is proportional to the SQUID-sample separation. (b) Temperature-dependent in-phase susceptibility of La$_{0.85}$Sr$_{0.15}$NiO$_2$ and Pr$_{0.8}$Sr$_{0.2}$NiO$_{2}$, obtained at the locations marked in (d) and (j), respectively. Each marker is determined from a fit such as those shown in (a). (c) Temperature history for the magnetometry scans of La$_{0.85}$Sr$_{0.15}$NiO$_2$ shown in (d)--(h). (i) Temperature history for the magnetometry scans of Pr$_{0.8}$Sr$_{0.2}$NiO$_{2}$ shown in (j) --(n). 
}
\label{main_temp}
\end{figure}

Given the presence of these ferromagnetic nanoparticles as a complicating factor in our films, we further investigated their magnetic behavior. Ferromagnetic nanoparticles are known to exhibit superparamagnetism, where the magnetic moments of single-domain nanoparticles appear to be frozen below a blocking temperature $T_B$ and collectively behave as a paramagnet above $T_B$ because of thermal activation \cite{superpara}. We observed similar blocking behaviors in local AC susceptibility and consecutive magnetometry scans, as shown in Fig.~\ref{main_temp}. We obtained the local magnetic susceptibility $\chi$, related to relative magnetic permeability by $\mu = \chi$ + 1, by fitting the susceptometry signal as a function of height. Techniques for determining magnetic properties, such as the magnetic permeability $\upmu$ and superfluid density $n_s$, by fitting the susceptometry signal vs height have been established in Ref. \cite{JohnTD}. Figure~\ref{main_temp}(a) displays data for La$_{0.85}$Sr$_{0.15}$NiO$_2$ at selected temperatures, fitted to a model of a uniform paramagnetic thin film \cite{JohnTD}. The magnetic susceptibility $\chi$ of La$_{0.85}$Sr$_{0.15}$NiO$_2$ and Pr$_{0.8}$Sr$_{0.2}$NiO$_2$ is fitted at each temperature, as summarized in Fig.~\ref{main_temp}(b). Figures~\ref{main_temp}(c)--\ref{main_temp}(h) and ~\ref{main_temp}(i)--\ref{main_temp}(n) display the temperature evolution of the ferromagnetic background in La$_{0.85}$Sr$_{0.15}$NiO$_2$ and Pr$_{0.8}$Sr$_{0.2}$NiO$_2$, respectively. The exact approach locations are indicated by markers in Figs.~\ref{main_temp}(d) and \ref{main_temp}(j). An ensemble of nanoparticles of uniform size will demonstrate a peak at the blocking temperature \cite{superpara}. For La$_{0.85}$Sr$_{0.15}$NiO$_2$, we observe an upturn from the base temperature to 17 K before the curve flattens, as shown in Fig.~\ref{main_temp}(b). The lack of a clear peak could be a result of a wide NiO$_x$ nanoparticle size distribution. In contrast, Pr$_{0.8}$Sr$_{0.2}$NiO$_2$ does not demonstrate a peak $\leq$ 33 K, indicating that the blocking temperature could be higher.  For La$_{0.85}$Sr$_{0.15}$NiO$_2$, increasing the temperature from 3 to 17 K does not significantly alter the ferromagnetic background, as exemplified in the upper-left coral box in Figs.~\ref{main_temp}(d)--\ref{main_temp}(f). However, increasing the temperature to 33 K, where the flattening in susceptibility has occurred, and then cooling back to the base temperature results in a fading of the boxed feature, as shown in Fig.~\ref{main_temp}(h). We also observed hysteresis, as highlighted by the purple box, where a magnetic feature persists upon thermal cycling to 33 K, showing a slightly altered shape after cooling to 3 K. Here, Pr$_{0.8}$Sr$_{0.2}$NiO$_2$, with a possible blocking temperature above the highest measured susceptibility at 33 K, demonstrates similar blocking behaviors in which the ferromagnetic background does not strongly deviate for thermal cycling $\leq$ 39 K, but is greatly modified when cycled to 125 K. This correlation between the ferromagnetic background in magnetometry and the seemingly superparamagnetic blocking susceptibility provides additional evidence that the ferromagnetic background is caused by magnetic nanoparticles.   
\par
Our results do not contradict previous works based on spectroscopic probes that demonstrated spin glass or short-range antiferromagnetism in Nd, Pr, and La nickelate thin films on STO ($\upmu$SR and RIXS) and powdered LaNiO$_2$ (NMR) \cite{muSR, powderSpinGlass, RIXS}. SQUID magnetometry cannot distinguish the existence or absence of antiferromagnetism, which has zero uncompensated spin except at edges and domain walls, nor spin glasses, which have microscopic randomly aligned spins, because both produce very small net magnetic flux. Searches for those orders would be better conducted with scattering probes or with scanning probes with finer sensitivity, better spatial resolution, and/or simultaneous measurement of sample topography, such as with probes such as spin-polarized scanning tunneling microscopy \cite{glass_STM} or magnetic force microscopy \cite{glass_AFM}. 
\par
The formation of NiO$_x$ nanoparticles due to synthesis imperfection could originate from two aspects: imperfect stoichiometry and lattice mismatch. Slight off-stoichiometry in the pulsed laser deposition plume could result in segregation of excess ingredients near the top sample surface in the form of nanoparticles. Lattice mismatch could also drive the formation of local defects to release the strain between the film and the substrate. One possible explanation is that the formation of nanoparticles relaxes the tensile epitaxial strain, especially during the growth of the initial perovskite $R$NiO$_3$ phase on STO. Based on these interpretations, the nanoparticle density could be reduced by selecting better lattice-matched substrates and more careful tuning of the synthesis stoichiometry.
\par

\begin{figure}[!htb]
\centering
  \includegraphics[scale = .45]{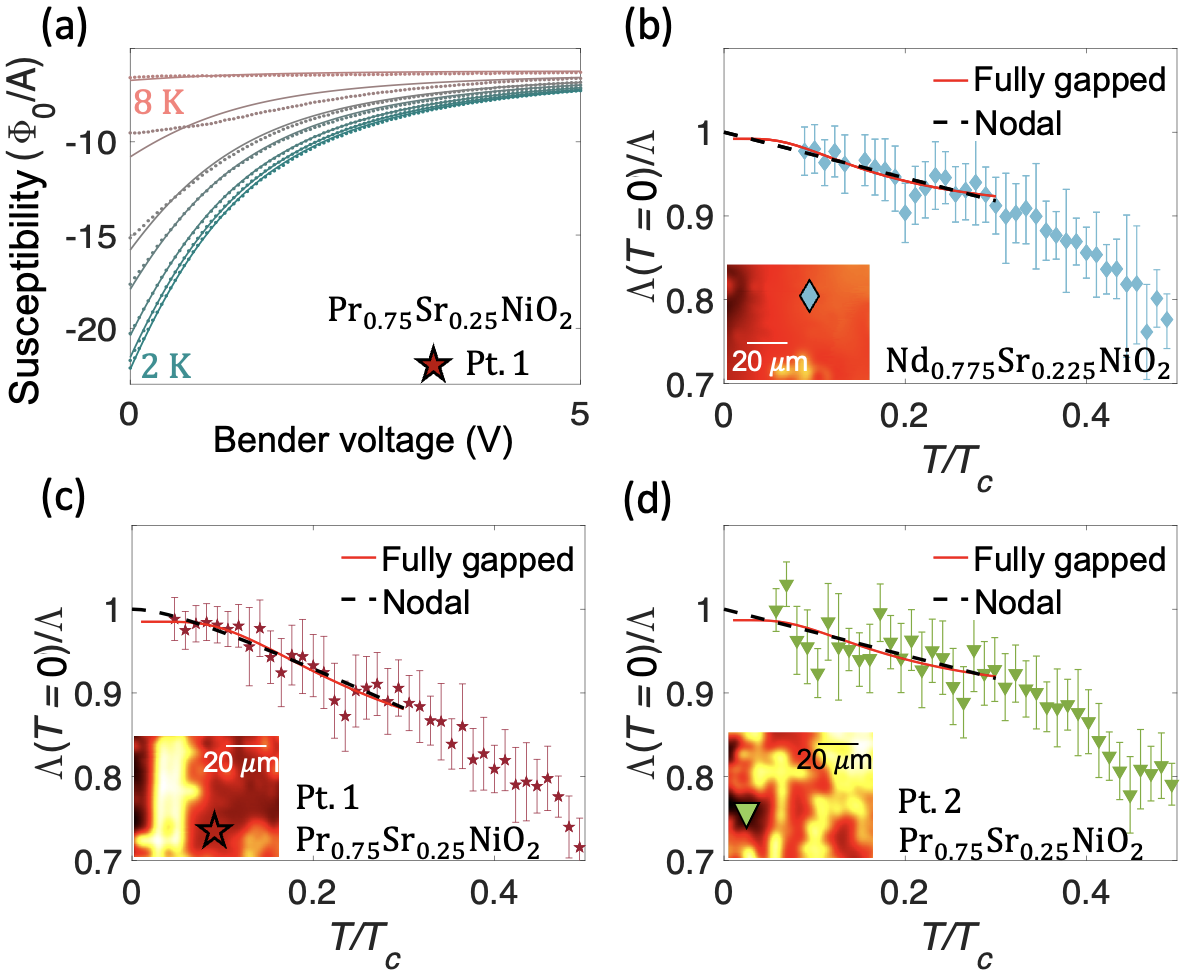}
  \caption{Temperature dependence of the superfluid density. (a) Susceptometry vs. bender voltage measured over Pr$_{0.8}$Sr$_{0.2}$NiO$_2$ at 2--8 K, shown with fits to a model assuming a diamagnetic thin film as described in the Supplemental Materials \cite{Supplemental}. These fits determine the Pearl length $\Lambda(T)$. Fitted values of $\Lambda(T=0)/\Lambda(T)$ vs normalized temperature determined at (b) the point in Nd$_{0.775}$Sr$_{0.225}$NiO$_2$ marked with a diamond in the inset, (c) the point in Pr$_{0.8}$Sr$_{0.2}$NiO$_2$ marked with a star in the inset, and (d) the point in Pr$_{0.8}$Sr$_{0.2}$NiO$_2$ marked with a triangle in the inset. The solid markers with error bars denote $1/\Lambda$ fitted from susceptibility approaches. The temperature dependence of $1/\Lambda$ was fitted to nodal (dashed lines) and non-nodal (solid lines) gap function models, with the best fits plotted. The zero-temperature Pearl length $\Lambda (T = 0)$ from the nodal fit was chosen as a normalization factor.  
}
\label{main_fit}
\end{figure}
We next discuss the characterization of the diamagnetic response. In thin-film superconductors, the Pearl length $\Lambda = 2\lambda^2/d$, where $\lambda$ and $d$ are the London penetration depth and film thickness, respectively, is the relevant length for characterizing diamagnetic screening \cite{PearlVortex}. The superfluid density and Pearl length are related by the equation $n_s = 2m^*/\mu_0\Lambda e^2d$, where $\mu_0$, $m^*$, and $e$ represent the vacuum permeability, effective charge carrier mass, and free electron charge, respectively \cite{SD_formula}. We fit our susceptibility vs height data to a thin-film diamagnetic model, as described in the Supplemental Material \cite{Supplemental} to extract the temperature-dependent superfluid density $n_s(T)$, denoted as $1/\Lambda(T)$ in the remainder of this paper\cite{JohnTD}. Our fitting model assumes a plane of uniform superfluid density. Therefore, we measure in relatively homogeneous regions to avoid the inhomogeneities seen in Figs.~\ref{main_coexist}(a) and (b). Figure~\ref{main_fit}(a) demonstrates satisfactory fitting at low temperatures, where the diamagnetic response of the superconductor is the main contribution to the susceptometry, and poor fitting near $T_c$, where the paramagnetic response [as shown in (Fig.~\ref{main_temp})] is larger and the diamagnetic response is smaller than at lower temperatures. Therefore, we only display fitted $\Lambda$ below 0.5 $T_c$ in Figs.~\ref{main_fit}(b)--\ref{main_fit}(d). All susceptibility approaches were made in regions that demonstrate relatively homogeneous diamagnetic screening over a range of tens of micrometers, as indicated on the inset of Fig.~3(b) - (d), in the hope that the neighboring inhomogeneities would not affect the measurement of the superfluid density.
\par
The temperature dependence of the superfluid density at low temperatures can indicate the presence of nodes in the superconducting gap. We measured the temperature dependence of the superfluid density $n_s(T)$ or the inverse Pearl length $1/\Lambda(T)$ at two locations on the Pr$_{0.8}$Sr$_{0.2}$NiO$_2$ film and one location on the Nd$_{0.775}$Sr$_{0.225}$NiO$_2$ film. In the Supplemental Material \cite{Supplemental}, we describe fitting to two models: a $d$-wave model with resonance scattering, in which the intermediate $T$-linear behavior crosses over to $T^2$ dependence below a crossover temperature $T^{**}$ owing to nonzero states near the Fermi surface \cite{CrossOver}, and an anisotropic $s$-wave model, in which $n_s$($T$) is exponential at temperatures that are small compared with the minimum gap $\Delta_0$, with a crossover to power-law-like behavior at intermediate temperatures. We also show that the fitted crossover temperature $T^{**}$ is close to or smaller than our lowest measurement temperature, such that our data can also be described by a $T$-linear $d$-wave model (see the Supplemental Material \cite{Supplemental}). In Figs.~\ref{main_fit}(b)--\ref{main_fit}(d), we display best fits to the model of a $d$-wave with scattering, labeled as nodal, and the anisotropic $s$-wave model, labeled as fully gapped. 
\par
The fitted coefficients in all three locations agree with macroscopic mutual-inductance measurements  (see the Supplemental Material \cite{Supplemental}) \cite{Hwang_SD} and are consistent with a $T$-linear temperature dependence from $\sim$0.1 $T_c$ to $\sim$0.3 $T_c$. One location measured on the Pr$_{0.8}$Sr$_{0.2}$NiO$_2$ film [Fig.~\ref{main_temp}(c)] may show a slight flattening, consistent with either a crossover to $T^2$ or the existence of a minimum gap, and we cannot rule out the existence of a crossover to $T^2$ in Nd$_{0.775}$Sr$_{0.225}$NiO$_2$, as reported in Ref.~\cite{Hwang_SD}. The fitted zero-temperature penetration depth $\lambda (T = 0)$, converted from the Pearl length $\Lambda (T = 0)$, is similar for both the nodal and fully gapped models for all measured regions. With the nominal assumption of 20 unit cells, $\lambda (T = 0)$ is in the range of 1.21--1.24 $\upmu$m for the two points in Pr$_{0.8}$Sr$_{0.2}$NiO$_2$, in excellent agreement with Ref. \cite{Hwang_SD} [$\lambda (T = 0)$ = 1.3 $\upmu$m]. Our data for the slightly more overdoped Nd$_{0.775}$Sr$_{0.225}$NiO$_2$ fit to both a nodal model and a fully gapped model, with a fitted $\lambda (T = 0)$ of $\sim$ 1.04 $\upmu$m in both models, like previous reports in optimally doped Nd$_{0.8}$Sr$_{0.2}$NiO$_2$ ($\lambda (T = 0)$= 0.75 $\upmu$m in Ref.~\cite{Hwang_SD} and $\lambda (T = 0)$= 0.58--1.46 $\upmu$m in Ref.~\cite{sing_SD}). With a predicted effective mass 2.8 times that of the free electron mass for Nd$_{0.8}$Sr$_{0.2}$NiO$_2$ \cite{DWaveJpn}, the zero-temperature superfluid density $n_s$ of our Nd$_{0.775}$Sr$_{0.225}$NiO$_2$ is estimated to be 7.33 $\times$ 10$^{25}$ /m$^3$. 
\par
We do not attempt to conclude $s$- or $d$-wave pairing symmetry from Figs.~\ref{main_fit}(b)--\ref{main_fit}(d). 
\par
Our results show that two forms of heterogeneity play a role that should be considered in fitting future measurements of $n_s$. First, the susceptibility in Figs.~\ref{main_temp}(a) and \ref{main_temp}(b) is dominated by the superparamagnetic susceptibility from NiO$_x$ inclusions. Although this signal decreases as the temperature decreases below the nominal blocking temperatures, it is still significant near $T_c$, where the diamagnetic susceptibility is small. Because this paramagnetic susceptibility resembles superparamagnetism with a broad distribution of blocking temperatures, neither the Curie-Weiss law nor any other specific functional form should be used to subtract this contribution. At lower temperatures, however, the paramagnetism decreases, and the measured diamagnetism is two orders of magnitude larger; thus, we have more confidence in our measured values of lower-temperature $n_s$. The error associated with the paramagnetism in the superfluid density fitting $leq$ 0.5 $T_c$ is estimated to be $\leq$ 2\%.  Second, the line- and cross-shaped features in the diamagnetic susceptibility [Figs.~\ref{main_coexist}(a) and \ref{main_coexist}(b)] complicate the measurement of $n_s$ at all temperatures; our relatively local measurement allows us to avoid those features to determine the low-temperature $n_s$.
\begin{figure}[htb]
\centering
  \includegraphics[scale = .55]{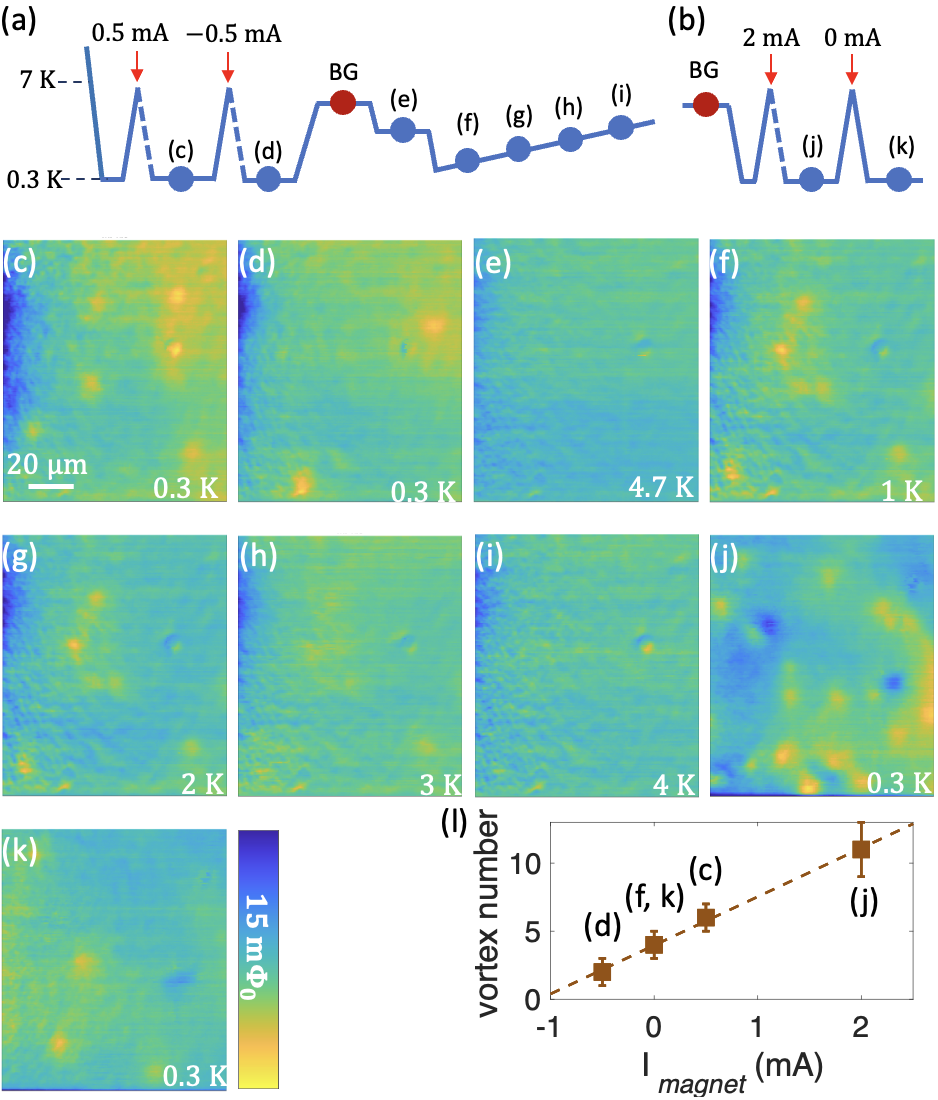}
  \caption{Superconducting vortices. Temperature and applied field history of two regions in Nd$_{0.775}$Sr$_{0.225}$NiO$_2$. The red dots are background scans, taken at (a) 6 K and (b) 4 K. The dashed lines indicate field cooling. (c)--(k) Background-subtracted magnetometry scans. The estimated scanning height is 4.6 $\upmu$m. The vortices appear as yellow regions representing a local peak in magnetic flux. (l) Number of vortices as a function of the cooling field, with corresponding scans labeled. The dashed line shows the best linear fit.}
\label{vortex}
\end{figure}

A natural question arises at this point: Where are the superconducting vortices? By subtracting images below the zero-resistance critical temperature $T_c$ (4.8 K for Nd$_{0.775}$Sr$_{0.225}$NiO$_2$) from a background scan taken above $T_c$, Fig.~\ref{vortex} reveals vortices in magnetometry images in two regions in Nd$_{0.775}$Sr$_{0.225}$NiO$_2$.  To control the number of vortices in each image, we applied a small magnetic field by running a current $I_{magnet}$ through a hand-made coil below the sample stage; the total magnetic field is the applied field, proportional to $I_{magnet}$, plus a stray background field. Figures~\ref{vortex}(a) and \ref{vortex}(b) present the temperature history of the sample for Figs.~\ref{vortex}(c)--\ref{vortex}(i) and \ref{vortex}(j) and (k), respectively.  As the temperature history in Fig. ~\ref{vortex}(a) shows, we first heated the sample to 7 K, which is above $T_c$ = 4.8 K. We then cooled in a field generated by $I_{magnet}$ = 0.5 mA before taking the data in Fig.~\ref{vortex}(c); reheated the sample to 7 K and then cooled with $I_{magnet}$ = -0.5 mA before taking the data in Fig.~\ref{vortex}(d); and heated the sample to 6 K to take an image of the ferromagnetic background, which is subtracted from the other images. Six pointlike features in Fig.~\ref{vortex}(c) and two pointlike features in Fig.~\ref{vortex}(d) are present. We identified these features as quantized superconducting vortices. We then cooled to 4.7 K in nominally zero field, $I_{magnet}$ = 0 mA, where we did not observe any vortices [Fig.~\ref{vortex}(e)]; further cooled to a temperature of 1 K, at which 4 vortices appeared at different locations [Fig.~\ref{vortex}(f)]; and then gradually warmed up, observing that the vortices remained at the same locations from 1 to 3 K and then disappeared at 4 K [Figs.~\ref{vortex}(f)--\ref{vortex}(i)]. We used a similar protocol in a different sample region [Figs.~\ref{vortex}(b),~\ref{vortex}(j), and~\ref{vortex}(k)] and found that cooling in an applied field of $I_{magnet}$ = 2-mA introduced 11 vortices that disappeared upon zero-field cooling (for details of vortex number counting, see Fig.~S13 in the Supplemental Material \cite{Supplemental}). 
\par
The disappearance of vortices near $T_c$ and the re-emergence upon field cooling at alternative locations meet our expectations. The vortex number counts as a function of the applied current through the magnet is shown in Fig. 4(l). These data are best fit by a linear slope of 3.57(0.12) /mA and an offset of 3.96 mA. Based on the geometry of the handmade magnet, we estimate that a 1~mA current should result in a change in the applied field of 6.7 mG at the film, causing a flux change of 72 $\upmu$m$^2$G and therefore applying 3.48 $\Phi_0$ in the scan. This slope is in excellent agreement with the experimental vortex number count. The fitted offset implies that the background field is 26.5 mG.

\begin{figure}[htb]
\centering
  \includegraphics[scale = .45]{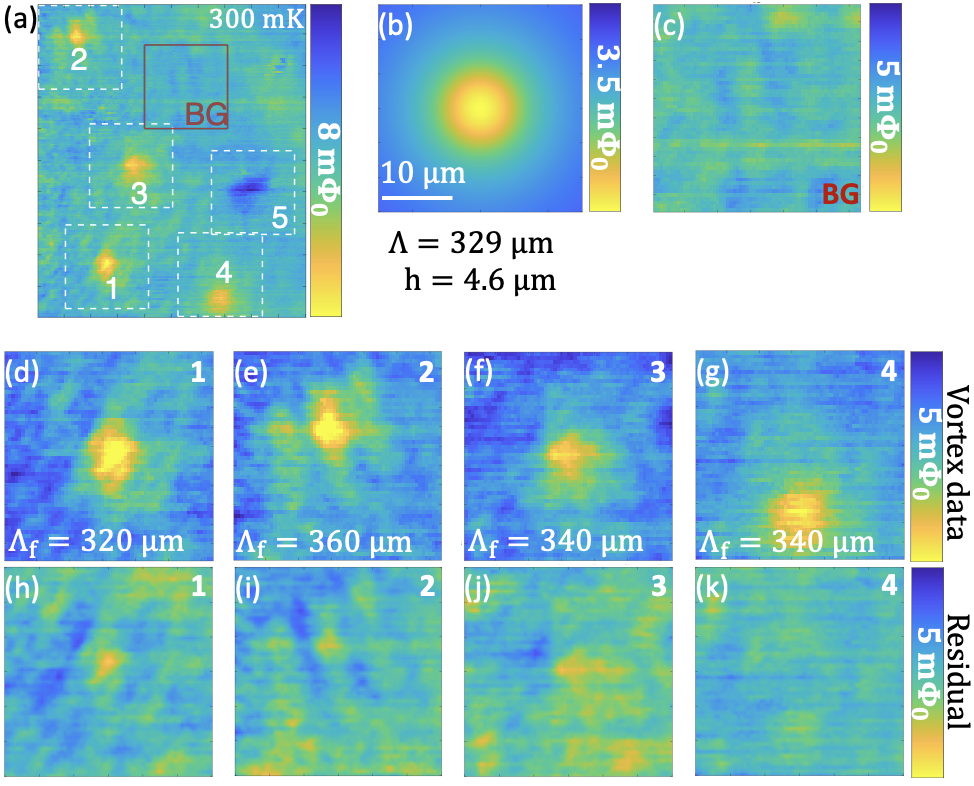}
  \caption{Close-up view of Pearl vortices in Nd$_{0.775}$Sr$_{0.225}$NiO$_2$ fitted to simulations. (a) Background-subtracted magnetometry scan from Fig.~\ref{vortex}(k). Five individual vortices (one of which is an antivortex) are boxed by dashed lines and labeled 1 -- 5. A random background acquired from a location away from the vortices is boxed by solid lines and labeled as BG. (b) Simulated Pearl vortex with a Pearl length of $\Lambda$ = 329 $\upmu$m at a scan height of $h$ = 4.6 $\upmu$m. (c) A region away from the vortices boxed by the red lines in (a) is taken as a random background. (d)--(g) Individual views of each vortex labeled in (a), with the fitted Pearl length $\Lambda_f$ indicated. The fitted scanning height $h_f$ from (d) to (g) is 4.8, 5, 5.2, and 5 $\upmu$m. (h)--(k) Residuals of individual vortex data subtracted from the simulation at the best fit. These residuals are similar in shape and strength to the random background in (c). }
\label{vortex_zoom}
\end{figure}

We determined the Pearl length by fitting the vortices in Fig.~\ref{vortex}(k) and compared the values with the Pearl length fitted from susceptibility approaches. The presence of a ferromagnetic background, particularly the residual background after an imperfect background subtraction, may alter the apparent shapes of the vortices. First, we subtracted a continuous background fitted to a third-degree polynomial plane from Fig.~\ref{vortex}(k) to better reveal vortices and an anti-vortex, as shown in Fig.~\ref{vortex_zoom}(a).  Figure~\ref{vortex_zoom}(b) displays a simulated Pearl vortex based on the parameters of a Pearl length of 329(26) $\upmu$m and a scan height of 4.6 $\upmu$m at 300 mK, as obtained from fitting the susceptibility approaches in this region. The vortices and the antivortex in Fig.~\ref{vortex_zoom}(a) sit in a background that originates from imperfect ferromagnetic background subtraction. An example of such a background is presented in Fig.~\ref{vortex_zoom}(c), with its location indicated by the solid box in Fig.~\ref{vortex_zoom}(a). The boxed features labeled 1--4 in Fig.~\ref{vortex_zoom}(a) are vortices and are shown in Figs.~\ref{vortex_zoom}(d)--\ref{vortex_zoom}(g). Box 5 shows an anti-vortex, which has a different sign than the vortices.  To extrapolate the Pearl length from vortex images, we fitted the vortices in Figs.~\ref{vortex_zoom}(d)--\ref{vortex_zoom}(g) by applying a two-parameter method to determine the most suitable scan height and Pearl length (see the Supplemental Material \cite{Supplemental}). The cross-sectional line of the Box 1 vortex fits is shown in Fig. S15(e) (see the Supplemental Material \cite{Supplemental}). The fitted Pearl lengths are indicated in Figs.~\ref{vortex_zoom}(d)--\ref{vortex_zoom}(g). The fitting quality is represented by the difference between each vortex and its best fit, denoted as the residual in Figs.~\ref{vortex_zoom}(h)--\ref{vortex_zoom}(k). The shape and magnitude of the residuals are comparable with the random background in Fig.~\ref{vortex_zoom}(c). The Pearl lengths obtained by fitting from susceptibility approaches and vortex images agree well. 
\par
Near the center of the Pearl vortex ($r<<\Lambda$), the out-of-plane magnetic field diverges as $1/r$, faster than an Abrikosov vortex which follows ln$(r/\lambda)$  \cite{PearlVortex}. In practice, this $1/r$ dependence means that the apparent width of the vortex is much smaller than the Pearl length. The full width half maximum of a typical Pearl vortex with a Pearl length of 350~$\upmu$m as seen by our SQUID at 4~$\upmu$m above can be estimated to be 8~$\upmu$m (see the Supplemental Material \cite{Supplemental}).
\par
In summary, we have conducted comprehensive scanning SQUID microscopy on superconducting infinite-layer nickelate films grown by pulsed laser deposition on STO substrates. First, we observed a ferromagnetic background [Figs.~\ref{main_coexist}(d)--(f)] whose behavior under thermal cycling is consistent with isolated superparamagnetic nanoparticles (Fig.~\ref{main_temp}). We attributed this magnetic signal to extrinsic NiO$_x$ nanoparticles whose chemical composition was identified by scanning transmission electron microscopy. The nanoparticle size and density were observed by atomic force microscopy, which allowed us to compare data to simulations. The dynamic susceptibility of such nanoparticles could complicate the interpretation of some bulk experiments for magnetic order and pairing symmetry. Second, we identified diamagnetic screening [Figs.~\ref{main_coexist}(b) and \ref{main_coexist}(d)] that appears inhomogeneous. Third, despite the ferromagnetic background and diamagnetic inhomogeneity, we observed a $T$-linear temperature dependence of our locally measured $n_s$ in Pr$_{0.8}$Sr$_{0.2}$NiO$_2$ and Nd$_{0.775}$Sr$_{0.225}$NiO$_2$, although our data do not go to sufficiently low temperature to conclude $s$- or $d$-wave pairing symmetry. Fourth, we found superconducting vortices and demonstrated that the number of these vortices is linear with respect to the applied magnetic field in the small-field regime. Finally, by analyzing the shapes of isolated vortices, we obtained a fitted Pearl length that agrees well with the Pearl length fitted from the AC susceptibility, with both methods yielding $\Lambda = 330 (30)\mu$m at 300 mK in a Nd$_{0.775}$Sr$_{0.225}$NiO$_2$ film with $T_c$ = 4.5 K. Our results highlight the potential importance of external NiO$_x$ magnetic inclusions in the interpretation of measurements on these and similar, and open research directions, such as studying vortex dynamics and flux quantization in superconducting nickelates. 
\par
This paper was supported by the Department of Energy, Office of Science, Basic Energy Sciences, Materials Sciences and Engineering Division, under contract No. DE-AC02-76SF00515. Sample synthesis was partially supported by the Gordon and Betty Moore Foundation’s Emergent Phenomena in Quantum Systems Initiative through Grant No. GBMF9072. B.H.G. and L.F.K. acknowledge support by the Department of Defense Air Force Office of Scientific Research (No. FA 9550-16-1-0305) and the Packard Foundation. The FEI Titan Themis 300 was acquired through support from the National Science Foundation (NSF; Grant No. NSF-MRI-1429155), with additional support from Cornell University, the Weill Institute, and the Kavli Institute at Cornell. The Thermo Fisher Helios G4 UX focused ion beam was acquired with support from NSF Grant No. DMR-2039380. 

\clearpage
\bibliography{main}

\begin{thebibliography}{56}%
\makeatletter
\providecommand \@ifxundefined [1]{%
 \@ifx{#1\undefined}
}%
\providecommand \@ifnum [1]{%
 \ifnum #1\expandafter \@firstoftwo
 \else \expandafter \@secondoftwo
 \fi
}%
\providecommand \@ifx [1]{%
 \ifx #1\expandafter \@firstoftwo
 \else \expandafter \@secondoftwo
 \fi
}%
\providecommand \natexlab [1]{#1}%
\providecommand \enquote  [1]{``#1''}%
\providecommand \bibnamefont  [1]{#1}%
\providecommand \bibfnamefont [1]{#1}%
\providecommand \citenamefont [1]{#1}%
\providecommand \href@noop [0]{\@secondoftwo}%
\providecommand \href [0]{\begingroup \@sanitize@url \@href}%
\providecommand \@href[1]{\@@startlink{#1}\@@href}%
\providecommand \@@href[1]{\endgroup#1\@@endlink}%
\providecommand \@sanitize@url [0]{\catcode `\\12\catcode `\$12\catcode
  `\&12\catcode `\#12\catcode `\^12\catcode `\_12\catcode `\%12\relax}%
\providecommand \@@startlink[1]{}%
\providecommand \@@endlink[0]{}%
\providecommand \url  [0]{\begingroup\@sanitize@url \@url }%
\providecommand \@url [1]{\endgroup\@href {#1}{\urlprefix }}%
\providecommand \urlprefix  [0]{URL }%
\providecommand \Eprint [0]{\href }%
\providecommand \doibase [0]{https://doi.org/}%
\providecommand \selectlanguage [0]{\@gobble}%
\providecommand \bibinfo  [0]{\@secondoftwo}%
\providecommand \bibfield  [0]{\@secondoftwo}%
\providecommand \translation [1]{[#1]}%
\providecommand \BibitemOpen [0]{}%
\providecommand \bibitemStop [0]{}%
\providecommand \bibitemNoStop [0]{.\EOS\space}%
\providecommand \EOS [0]{\spacefactor3000\relax}%
\providecommand \BibitemShut  [1]{\csname bibitem#1\endcsname}%
\let\auto@bib@innerbib\@empty
\bibitem [{\citenamefont {Li}\ \emph {et~al.}(2019)\citenamefont {Li},
  \citenamefont {Lee}, \citenamefont {Wang}, \citenamefont {Osada},
  \citenamefont {Crossley}, \citenamefont {Lee}, \citenamefont {Cui},
  \citenamefont {Hikita},\ and\ \citenamefont {Hwang}}]{discovery}%
  \BibitemOpen
  \bibfield  {author} {\bibinfo {author} {\bibfnamefont {D.}~\bibnamefont
  {Li}}, \bibinfo {author} {\bibfnamefont {K.}~\bibnamefont {Lee}}, \bibinfo
  {author} {\bibfnamefont {B.~Y.}\ \bibnamefont {Wang}}, \bibinfo {author}
  {\bibfnamefont {M.}~\bibnamefont {Osada}}, \bibinfo {author} {\bibfnamefont
  {S.}~\bibnamefont {Crossley}}, \bibinfo {author} {\bibfnamefont {H.~R.}\
  \bibnamefont {Lee}}, \bibinfo {author} {\bibfnamefont {Y.}~\bibnamefont
  {Cui}}, \bibinfo {author} {\bibfnamefont {Y.}~\bibnamefont {Hikita}},\ and\
  \bibinfo {author} {\bibfnamefont {H.~Y.}\ \bibnamefont {Hwang}},\ }\bibfield
  {title} {\bibinfo {title} {Superconductivity in an infinite-layer
  nickelate},\ }\href {https://doi.org/10.1038/s41586-019-1496-5} {\bibfield
  {journal} {\bibinfo  {journal} {Nature}\ }\textbf {\bibinfo {volume} {572}},\
  \bibinfo {pages} {624} (\bibinfo {year} {2019})}\BibitemShut {NoStop}%
\bibitem [{\citenamefont {Botana}\ \emph {et~al.}(2021)\citenamefont {Botana},
  \citenamefont {Bernardini},\ and\ \citenamefont {Cano}}]{Botana_2021}%
  \BibitemOpen
  \bibfield  {author} {\bibinfo {author} {\bibfnamefont {A.~S.}\ \bibnamefont
  {Botana}}, \bibinfo {author} {\bibfnamefont {F.}~\bibnamefont {Bernardini}},\
  and\ \bibinfo {author} {\bibfnamefont {A.}~\bibnamefont {Cano}},\ }\bibfield
  {title} {\bibinfo {title} {Nickelate superconductors: An ongoing dialog
  between theory and experiments},\ }\href
  {https://doi.org/10.1134/s1063776121040026} {\bibfield  {journal} {\bibinfo
  {journal} {Journal of Experimental and Theoretical Physics}\ }\textbf
  {\bibinfo {volume} {132}},\ \bibinfo {pages} {618} (\bibinfo {year}
  {2021})}\BibitemShut {NoStop}%
\bibitem [{\citenamefont {Zhou}\ \emph {et~al.}(2022)\citenamefont {Zhou},
  \citenamefont {Qin}, \citenamefont {Feng}, \citenamefont {Yan}, \citenamefont
  {Wang}, \citenamefont {Chen}, \citenamefont {Meng},\ and\ \citenamefont
  {Liu}}]{ChineseRev}%
  \BibitemOpen
  \bibfield  {author} {\bibinfo {author} {\bibfnamefont {X.}~\bibnamefont
  {Zhou}}, \bibinfo {author} {\bibfnamefont {P.}~\bibnamefont {Qin}}, \bibinfo
  {author} {\bibfnamefont {Z.}~\bibnamefont {Feng}}, \bibinfo {author}
  {\bibfnamefont {H.}~\bibnamefont {Yan}}, \bibinfo {author} {\bibfnamefont
  {X.}~\bibnamefont {Wang}}, \bibinfo {author} {\bibfnamefont {H.}~\bibnamefont
  {Chen}}, \bibinfo {author} {\bibfnamefont {Z.}~\bibnamefont {Meng}},\ and\
  \bibinfo {author} {\bibfnamefont {Z.}~\bibnamefont {Liu}},\ }\bibfield
  {title} {\bibinfo {title} {Experimental progress on the emergent
  infinite-layer {Ni-based} superconductors},\ }\href
  {https://doi.org/10.1016/j.mattod.2022.02.016} {\bibfield  {journal}
  {\bibinfo  {journal} {Materials Today}\ }\textbf {\bibinfo {volume} {55}},\
  \bibinfo {pages} {170} (\bibinfo {year} {2022})}\BibitemShut {NoStop}%
\bibitem [{\citenamefont {Nomura}\ and\ \citenamefont
  {Arita}(2022)}]{JapaneseRev}%
  \BibitemOpen
  \bibfield  {author} {\bibinfo {author} {\bibfnamefont {Y.}~\bibnamefont
  {Nomura}}\ and\ \bibinfo {author} {\bibfnamefont {R.}~\bibnamefont {Arita}},\
  }\bibfield  {title} {\bibinfo {title} {Superconductivity in infinite-layer
  nickelates},\ }\href {https://doi.org/10.1088/1361-6633/ac5a60} {\bibfield
  {journal} {\bibinfo  {journal} {Reports on Progress in Physics}\ }\textbf
  {\bibinfo {volume} {85}},\ \bibinfo {pages} {052501} (\bibinfo {year}
  {2022})}\BibitemShut {NoStop}%
\bibitem [{\citenamefont {Yang}\ and\ \citenamefont
  {Zhang}(2022)}]{FrontierYang}%
  \BibitemOpen
  \bibfield  {author} {\bibinfo {author} {\bibfnamefont {Y.-f.}\ \bibnamefont
  {Yang}}\ and\ \bibinfo {author} {\bibfnamefont {G.-M.}\ \bibnamefont
  {Zhang}},\ }\bibfield  {title} {\bibinfo {title} {Self-doping and the
  {Mott-Kondo} scenario for infinite-layer nickelate superconductors},\ }\href
  {https://doi.org/10.3389/fphy.2021.801236} {\bibfield  {journal} {\bibinfo
  {journal} {Frontiers in Physics}\ }\textbf {\bibinfo {volume} {9}},\ \bibinfo
  {pages} {783} (\bibinfo {year} {2022})}\BibitemShut {NoStop}%
\bibitem [{\citenamefont {Mitchell}(2021)}]{FrontierMitchell}%
  \BibitemOpen
  \bibfield  {author} {\bibinfo {author} {\bibfnamefont {J.~F.}\ \bibnamefont
  {Mitchell}},\ }\bibfield  {title} {\bibinfo {title} {A nickelate
  renaissance},\ }\href {https://doi.org/10.3389/fphy.2021.813483} {\bibfield
  {journal} {\bibinfo  {journal} {Frontiers in Physics}\ }\textbf {\bibinfo
  {volume} {9}},\ \bibinfo {pages} {813483} (\bibinfo {year}
  {2021})}\BibitemShut {NoStop}%
\bibitem [{\citenamefont {Botana}\ \emph {et~al.}(2022)\citenamefont {Botana},
  \citenamefont {Lee}, \citenamefont {Norman}, \citenamefont {Pardo},\ and\
  \citenamefont {Pickett}}]{FrontierBotana}%
  \BibitemOpen
  \bibfield  {author} {\bibinfo {author} {\bibfnamefont {A.~S.}\ \bibnamefont
  {Botana}}, \bibinfo {author} {\bibfnamefont {K.-W.}\ \bibnamefont {Lee}},
  \bibinfo {author} {\bibfnamefont {M.~R.}\ \bibnamefont {Norman}}, \bibinfo
  {author} {\bibfnamefont {V.}~\bibnamefont {Pardo}},\ and\ \bibinfo {author}
  {\bibfnamefont {W.~E.}\ \bibnamefont {Pickett}},\ }\bibfield  {title}
  {\bibinfo {title} {Low valence nickelates: Launching the nickel age of
  superconductivity},\ }\href {https://doi.org/10.3389/fphy.2021.813532}
  {\bibfield  {journal} {\bibinfo  {journal} {Frontiers in Physics}\ }\textbf
  {\bibinfo {volume} {9}},\ \bibinfo {pages} {812} (\bibinfo {year}
  {2022})}\BibitemShut {NoStop}%
\bibitem [{\citenamefont {Li}\ \emph {et~al.}(2020)\citenamefont {Li},
  \citenamefont {Wang}, \citenamefont {Lee}, \citenamefont {Harvey},
  \citenamefont {Osada}, \citenamefont {Goodge}, \citenamefont {Kourkoutis},\
  and\ \citenamefont {Hwang}}]{SrDome}%
  \BibitemOpen
  \bibfield  {author} {\bibinfo {author} {\bibfnamefont {D.}~\bibnamefont
  {Li}}, \bibinfo {author} {\bibfnamefont {B.~Y.}\ \bibnamefont {Wang}},
  \bibinfo {author} {\bibfnamefont {K.}~\bibnamefont {Lee}}, \bibinfo {author}
  {\bibfnamefont {S.~P.}\ \bibnamefont {Harvey}}, \bibinfo {author}
  {\bibfnamefont {M.}~\bibnamefont {Osada}}, \bibinfo {author} {\bibfnamefont
  {B.~H.}\ \bibnamefont {Goodge}}, \bibinfo {author} {\bibfnamefont {L.~F.}\
  \bibnamefont {Kourkoutis}},\ and\ \bibinfo {author} {\bibfnamefont {H.~Y.}\
  \bibnamefont {Hwang}},\ }\bibfield  {title} {\bibinfo {title}
  {Superconducting dome in
  {${\mathrm{Nd}}_{1\ensuremath{-}x}{\mathrm{Sr}}_{x}{\mathrm{NiO}}_{2}$}
  infinite layer films},\ }\href
  {https://doi.org/10.1103/PhysRevLett.125.027001} {\bibfield  {journal}
  {\bibinfo  {journal} {Phys. Rev. Lett.}\ }\textbf {\bibinfo {volume} {125}},\
  \bibinfo {pages} {027001} (\bibinfo {year} {2020})}\BibitemShut {NoStop}%
\bibitem [{\citenamefont {Osada}\ \emph {et~al.}(2020)\citenamefont {Osada},
  \citenamefont {Wang}, \citenamefont {Goodge}, \citenamefont {Lee},
  \citenamefont {Yoon}, \citenamefont {Sakuma}, \citenamefont {Li},
  \citenamefont {Miura}, \citenamefont {Kourkoutis},\ and\ \citenamefont
  {Hwang}}]{PrDome}%
  \BibitemOpen
  \bibfield  {author} {\bibinfo {author} {\bibfnamefont {M.}~\bibnamefont
  {Osada}}, \bibinfo {author} {\bibfnamefont {B.~Y.}\ \bibnamefont {Wang}},
  \bibinfo {author} {\bibfnamefont {B.~H.}\ \bibnamefont {Goodge}}, \bibinfo
  {author} {\bibfnamefont {K.}~\bibnamefont {Lee}}, \bibinfo {author}
  {\bibfnamefont {H.}~\bibnamefont {Yoon}}, \bibinfo {author} {\bibfnamefont
  {K.}~\bibnamefont {Sakuma}}, \bibinfo {author} {\bibfnamefont
  {D.}~\bibnamefont {Li}}, \bibinfo {author} {\bibfnamefont {M.}~\bibnamefont
  {Miura}}, \bibinfo {author} {\bibfnamefont {L.~F.}\ \bibnamefont
  {Kourkoutis}},\ and\ \bibinfo {author} {\bibfnamefont {H.~Y.}\ \bibnamefont
  {Hwang}},\ }\bibfield  {title} {\bibinfo {title} {A superconducting
  praseodymium nickelate with infinite layer structure},\ }\href
  {https://doi.org/10.1021/acs.nanolett.0c01392} {\bibfield  {journal}
  {\bibinfo  {journal} {Nano Letters}\ }\textbf {\bibinfo {volume} {20}},\
  \bibinfo {pages} {5735} (\bibinfo {year} {2020})}\BibitemShut {NoStop}%
\bibitem [{\citenamefont {Goodge}\ \emph {et~al.}(2021)\citenamefont {Goodge},
  \citenamefont {Li}, \citenamefont {Lee}, \citenamefont {Osada}, \citenamefont
  {Wang}, \citenamefont {Sawatzky}, \citenamefont {Hwang},\ and\ \citenamefont
  {Kourkoutis}}]{Goodge}%
  \BibitemOpen
  \bibfield  {author} {\bibinfo {author} {\bibfnamefont {B.~H.}\ \bibnamefont
  {Goodge}}, \bibinfo {author} {\bibfnamefont {D.}~\bibnamefont {Li}}, \bibinfo
  {author} {\bibfnamefont {K.}~\bibnamefont {Lee}}, \bibinfo {author}
  {\bibfnamefont {M.}~\bibnamefont {Osada}}, \bibinfo {author} {\bibfnamefont
  {B.~Y.}\ \bibnamefont {Wang}}, \bibinfo {author} {\bibfnamefont {G.~A.}\
  \bibnamefont {Sawatzky}}, \bibinfo {author} {\bibfnamefont {H.~Y.}\
  \bibnamefont {Hwang}},\ and\ \bibinfo {author} {\bibfnamefont {L.~F.}\
  \bibnamefont {Kourkoutis}},\ }\bibfield  {title} {\bibinfo {title} {Doping
  evolution of the {Mott–Hubbard} landscape in infinite-layer nickelates},\
  }\href {https://doi.org/10.1073/pnas.2007683118} {\bibfield  {journal}
  {\bibinfo  {journal} {Proceedings of the National Academy of Sciences}\
  }\textbf {\bibinfo {volume} {118}},\ \bibinfo {pages} {e2007683118} (\bibinfo
  {year} {2021})}\BibitemShut {NoStop}%
\bibitem [{\citenamefont {Hepting}\ \emph {et~al.}(2020)\citenamefont
  {Hepting}, \citenamefont {Li}, \citenamefont {Jia}, \citenamefont {Lu},
  \citenamefont {Paris}, \citenamefont {Tseng}, \citenamefont {Feng},
  \citenamefont {Osada}, \citenamefont {Been}, \citenamefont {Hikita},
  \citenamefont {Chuang}, \citenamefont {Hussain}, \citenamefont {Zhou},
  \citenamefont {Nag}, \citenamefont {Garcia-Fernandez}, \citenamefont {Rossi},
  \citenamefont {Huang}, \citenamefont {Huang}, \citenamefont {Shen},
  \citenamefont {Schmitt}, \citenamefont {Hwang}, \citenamefont {Moritz},
  \citenamefont {Zaanen}, \citenamefont {Devereaux},\ and\ \citenamefont
  {Lee}}]{XRay}%
  \BibitemOpen
  \bibfield  {author} {\bibinfo {author} {\bibfnamefont {M.}~\bibnamefont
  {Hepting}}, \bibinfo {author} {\bibfnamefont {D.}~\bibnamefont {Li}},
  \bibinfo {author} {\bibfnamefont {C.~J.}\ \bibnamefont {Jia}}, \bibinfo
  {author} {\bibfnamefont {H.}~\bibnamefont {Lu}}, \bibinfo {author}
  {\bibfnamefont {E.}~\bibnamefont {Paris}}, \bibinfo {author} {\bibfnamefont
  {Y.}~\bibnamefont {Tseng}}, \bibinfo {author} {\bibfnamefont
  {X.}~\bibnamefont {Feng}}, \bibinfo {author} {\bibfnamefont {M.}~\bibnamefont
  {Osada}}, \bibinfo {author} {\bibfnamefont {E.}~\bibnamefont {Been}},
  \bibinfo {author} {\bibfnamefont {Y.}~\bibnamefont {Hikita}}, \bibinfo
  {author} {\bibfnamefont {Y.-D.}\ \bibnamefont {Chuang}}, \bibinfo {author}
  {\bibfnamefont {Z.}~\bibnamefont {Hussain}}, \bibinfo {author} {\bibfnamefont
  {K.~J.}\ \bibnamefont {Zhou}}, \bibinfo {author} {\bibfnamefont
  {A.}~\bibnamefont {Nag}}, \bibinfo {author} {\bibfnamefont {M.}~\bibnamefont
  {Garcia-Fernandez}}, \bibinfo {author} {\bibfnamefont {M.}~\bibnamefont
  {Rossi}}, \bibinfo {author} {\bibfnamefont {H.~Y.}\ \bibnamefont {Huang}},
  \bibinfo {author} {\bibfnamefont {D.~J.}\ \bibnamefont {Huang}}, \bibinfo
  {author} {\bibfnamefont {Z.~X.}\ \bibnamefont {Shen}}, \bibinfo {author}
  {\bibfnamefont {T.}~\bibnamefont {Schmitt}}, \bibinfo {author} {\bibfnamefont
  {H.~Y.}\ \bibnamefont {Hwang}}, \bibinfo {author} {\bibfnamefont
  {B.}~\bibnamefont {Moritz}}, \bibinfo {author} {\bibfnamefont
  {J.}~\bibnamefont {Zaanen}}, \bibinfo {author} {\bibfnamefont {T.~P.}\
  \bibnamefont {Devereaux}},\ and\ \bibinfo {author} {\bibfnamefont {W.~S.}\
  \bibnamefont {Lee}},\ }\bibfield  {title} {\bibinfo {title} {Electronic
  structure of the parent compound of superconducting infinite-layer
  nickelates},\ }\href {https://doi.org/10.1038/s41563-019-0585-z} {\bibfield
  {journal} {\bibinfo  {journal} {Nature Materials}\ }\textbf {\bibinfo
  {volume} {19}},\ \bibinfo {pages} {381} (\bibinfo {year} {2020})}\BibitemShut
  {NoStop}%
\bibitem [{\citenamefont {Chen}\ \emph
  {et~al.}(2022{\natexlab{a}})\citenamefont {Chen}, \citenamefont {Osada},
  \citenamefont {Li}, \citenamefont {Been}, \citenamefont {Chen}, \citenamefont
  {Hashimoto}, \citenamefont {Lu}, \citenamefont {Mo}, \citenamefont {Lee},
  \citenamefont {Wang}, \citenamefont {Rodolakis}, \citenamefont {McChesney},
  \citenamefont {Jia}, \citenamefont {Moritz}, \citenamefont {Devereaux},
  \citenamefont {Hwang},\ and\ \citenamefont {Shen}}]{ARPES}%
  \BibitemOpen
  \bibfield  {author} {\bibinfo {author} {\bibfnamefont {Z.}~\bibnamefont
  {Chen}}, \bibinfo {author} {\bibfnamefont {M.}~\bibnamefont {Osada}},
  \bibinfo {author} {\bibfnamefont {D.}~\bibnamefont {Li}}, \bibinfo {author}
  {\bibfnamefont {E.~M.}\ \bibnamefont {Been}}, \bibinfo {author}
  {\bibfnamefont {S.-D.}\ \bibnamefont {Chen}}, \bibinfo {author}
  {\bibfnamefont {M.}~\bibnamefont {Hashimoto}}, \bibinfo {author}
  {\bibfnamefont {D.}~\bibnamefont {Lu}}, \bibinfo {author} {\bibfnamefont
  {S.-K.}\ \bibnamefont {Mo}}, \bibinfo {author} {\bibfnamefont
  {K.}~\bibnamefont {Lee}}, \bibinfo {author} {\bibfnamefont {B.~Y.}\
  \bibnamefont {Wang}}, \bibinfo {author} {\bibfnamefont {F.}~\bibnamefont
  {Rodolakis}}, \bibinfo {author} {\bibfnamefont {J.~L.}\ \bibnamefont
  {McChesney}}, \bibinfo {author} {\bibfnamefont {C.}~\bibnamefont {Jia}},
  \bibinfo {author} {\bibfnamefont {B.}~\bibnamefont {Moritz}}, \bibinfo
  {author} {\bibfnamefont {T.~P.}\ \bibnamefont {Devereaux}}, \bibinfo {author}
  {\bibfnamefont {H.~Y.}\ \bibnamefont {Hwang}},\ and\ \bibinfo {author}
  {\bibfnamefont {Z.-X.}\ \bibnamefont {Shen}},\ }\bibfield  {title} {\bibinfo
  {title} {Electronic structure of superconducting nickelates probed by
  resonant photoemission spectroscopy},\ }\href
  {https://doi.org/https://doi.org/10.1016/j.matt.2022.01.020} {\bibfield
  {journal} {\bibinfo  {journal} {Matter}\ }\textbf {\bibinfo {volume} {5}},\
  \bibinfo {pages} {1806} (\bibinfo {year} {2022}{\natexlab{a}})}\BibitemShut
  {NoStop}%
\bibitem [{\citenamefont {Held}\ \emph {et~al.}(2022)\citenamefont {Held},
  \citenamefont {Si}, \citenamefont {Worm}, \citenamefont {Janson},
  \citenamefont {Arita}, \citenamefont {Zhong}, \citenamefont {Tomczak},\ and\
  \citenamefont {Kitatani}}]{FrontierHeld}%
  \BibitemOpen
  \bibfield  {author} {\bibinfo {author} {\bibfnamefont {K.}~\bibnamefont
  {Held}}, \bibinfo {author} {\bibfnamefont {L.}~\bibnamefont {Si}}, \bibinfo
  {author} {\bibfnamefont {P.}~\bibnamefont {Worm}}, \bibinfo {author}
  {\bibfnamefont {O.}~\bibnamefont {Janson}}, \bibinfo {author} {\bibfnamefont
  {R.}~\bibnamefont {Arita}}, \bibinfo {author} {\bibfnamefont
  {Z.}~\bibnamefont {Zhong}}, \bibinfo {author} {\bibfnamefont {J.~M.}\
  \bibnamefont {Tomczak}},\ and\ \bibinfo {author} {\bibfnamefont
  {M.}~\bibnamefont {Kitatani}},\ }\bibfield  {title} {\bibinfo {title} {Phase
  diagram of nickelate superconductors calculated by dynamical vertex
  approximation},\ }\href {https://doi.org/10.3389/fphy.2021.810394} {\bibfield
   {journal} {\bibinfo  {journal} {Frontiers in Physics}\ }\textbf {\bibinfo
  {volume} {9}},\ \bibinfo {pages} {803} (\bibinfo {year} {2022})}\BibitemShut
  {NoStop}%
\bibitem [{\citenamefont {Hirayama}\ \emph {et~al.}(2022)\citenamefont
  {Hirayama}, \citenamefont {Nomura},\ and\ \citenamefont
  {Arita}}]{FrontierHirayama}%
  \BibitemOpen
  \bibfield  {author} {\bibinfo {author} {\bibfnamefont {M.}~\bibnamefont
  {Hirayama}}, \bibinfo {author} {\bibfnamefont {Y.}~\bibnamefont {Nomura}},\
  and\ \bibinfo {author} {\bibfnamefont {R.}~\bibnamefont {Arita}},\ }\bibfield
   {title} {\bibinfo {title} {Ab initio downfolding based on the {GW}
  approximation for infinite-layer nickelates},\ }\href
  {https://doi.org/10.3389/fphy.2022.824144} {\bibfield  {journal} {\bibinfo
  {journal} {Frontiers in Physics}\ }\textbf {\bibinfo {volume} {10}},\
  \bibinfo {pages} {49} (\bibinfo {year} {2022})}\BibitemShut {NoStop}%
\bibitem [{\citenamefont {Jiang}\ \emph {et~al.}(2020)\citenamefont {Jiang},
  \citenamefont {Berciu},\ and\ \citenamefont {Sawatzky}}]{AFM10meV}%
  \BibitemOpen
  \bibfield  {author} {\bibinfo {author} {\bibfnamefont {M.}~\bibnamefont
  {Jiang}}, \bibinfo {author} {\bibfnamefont {M.}~\bibnamefont {Berciu}},\ and\
  \bibinfo {author} {\bibfnamefont {G.~A.}\ \bibnamefont {Sawatzky}},\
  }\bibfield  {title} {\bibinfo {title} {Critical nature of the {Ni} spin state
  in doped {NdNiO$_2$}},\ }\href
  {https://doi.org/10.1103/PhysRevLett.124.207004} {\bibfield  {journal}
  {\bibinfo  {journal} {Phys. Rev. Lett.}\ }\textbf {\bibinfo {volume} {124}},\
  \bibinfo {pages} {207004} (\bibinfo {year} {2020})}\BibitemShut {NoStop}%
\bibitem [{\citenamefont {Liu}\ \emph {et~al.}(2020)\citenamefont {Liu},
  \citenamefont {Ren}, \citenamefont {Zhu}, \citenamefont {Wang},\ and\
  \citenamefont {Yang}}]{2020AFMtraces}%
  \BibitemOpen
  \bibfield  {author} {\bibinfo {author} {\bibfnamefont {Z.}~\bibnamefont
  {Liu}}, \bibinfo {author} {\bibfnamefont {Z.}~\bibnamefont {Ren}}, \bibinfo
  {author} {\bibfnamefont {W.}~\bibnamefont {Zhu}}, \bibinfo {author}
  {\bibfnamefont {Z.}~\bibnamefont {Wang}},\ and\ \bibinfo {author}
  {\bibfnamefont {J.}~\bibnamefont {Yang}},\ }\bibfield  {title} {\bibinfo
  {title} {Electronic and magnetic structure of infinite-layer {NdNiO}$_2$:
  trace of antiferromagnetic metal},\ }\href
  {https://doi.org/10.1038/s41535-020-0229-1} {\bibfield  {journal} {\bibinfo
  {journal} {npj Quantum Materials}\ }\textbf {\bibinfo {volume} {5}},\
  \bibinfo {pages} {31} (\bibinfo {year} {2020})}\BibitemShut {NoStop}%
\bibitem [{\citenamefont {Kapeghian}\ and\ \citenamefont
  {Botana}(2020)}]{2020AFMUniversal}%
  \BibitemOpen
  \bibfield  {author} {\bibinfo {author} {\bibfnamefont {J.}~\bibnamefont
  {Kapeghian}}\ and\ \bibinfo {author} {\bibfnamefont {A.~S.}\ \bibnamefont
  {Botana}},\ }\bibfield  {title} {\bibinfo {title} {Electronic structure and
  magnetism in infinite-layer nickelates {RNiO}$_2$
  ({R}=$\mathrm{La}\text{\ensuremath{-}}\mathrm{Lu}$)},\ }\href
  {https://doi.org/10.1103/PhysRevB.102.205130} {\bibfield  {journal} {\bibinfo
   {journal} {Phys. Rev. B}\ }\textbf {\bibinfo {volume} {102}},\ \bibinfo
  {pages} {205130} (\bibinfo {year} {2020})}\BibitemShut {NoStop}%
\bibitem [{\citenamefont {Zhang}\ \emph {et~al.}(2021)\citenamefont {Zhang},
  \citenamefont {Lane}, \citenamefont {Singh}, \citenamefont {Nokelainen},
  \citenamefont {Barbiellini}, \citenamefont {Markiewicz}, \citenamefont
  {Bansil},\ and\ \citenamefont {Sun}}]{2021AFM}%
  \BibitemOpen
  \bibfield  {author} {\bibinfo {author} {\bibfnamefont {R.}~\bibnamefont
  {Zhang}}, \bibinfo {author} {\bibfnamefont {C.}~\bibnamefont {Lane}},
  \bibinfo {author} {\bibfnamefont {B.}~\bibnamefont {Singh}}, \bibinfo
  {author} {\bibfnamefont {J.}~\bibnamefont {Nokelainen}}, \bibinfo {author}
  {\bibfnamefont {B.}~\bibnamefont {Barbiellini}}, \bibinfo {author}
  {\bibfnamefont {R.~S.}\ \bibnamefont {Markiewicz}}, \bibinfo {author}
  {\bibfnamefont {A.}~\bibnamefont {Bansil}},\ and\ \bibinfo {author}
  {\bibfnamefont {J.}~\bibnamefont {Sun}},\ }\bibfield  {title} {\bibinfo
  {title} {Magnetic and f-electron effects in {LaNiO$_2$} and {NdNiO$_2$}
  nickelates with cuprate-like 3$d_{x^2 - y^2}$ band},\ }\href
  {https://doi.org/10.1038/s42005-021-00621-4} {\bibfield  {journal} {\bibinfo
  {journal} {Communications Physics}\ }\textbf {\bibinfo {volume} {4}},\
  \bibinfo {pages} {118} (\bibinfo {year} {2021})}\BibitemShut {NoStop}%
\bibitem [{\citenamefont {Chen}\ \emph
  {et~al.}(2022{\natexlab{b}})\citenamefont {Chen}, \citenamefont {Hampel},
  \citenamefont {Karp}, \citenamefont {Lechermann},\ and\ \citenamefont
  {Millis}}]{FrontierChen}%
  \BibitemOpen
  \bibfield  {author} {\bibinfo {author} {\bibfnamefont {H.}~\bibnamefont
  {Chen}}, \bibinfo {author} {\bibfnamefont {A.}~\bibnamefont {Hampel}},
  \bibinfo {author} {\bibfnamefont {J.}~\bibnamefont {Karp}}, \bibinfo {author}
  {\bibfnamefont {F.}~\bibnamefont {Lechermann}},\ and\ \bibinfo {author}
  {\bibfnamefont {A.~J.}\ \bibnamefont {Millis}},\ }\bibfield  {title}
  {\bibinfo {title} {Dynamical mean field studies of infinite layer nickelates:
  Physics results and methodological implications},\ }\href
  {https://doi.org/10.3389/fphy.2022.835942} {\bibfield  {journal} {\bibinfo
  {journal} {Frontiers in Physics}\ }\textbf {\bibinfo {volume} {10}},\
  \bibinfo {pages} {835942} (\bibinfo {year} {2022}{\natexlab{b}})}\BibitemShut
  {NoStop}%
\bibitem [{\citenamefont {Wan}\ \emph {et~al.}(2021)\citenamefont {Wan},
  \citenamefont {Ivanov}, \citenamefont {Resta}, \citenamefont {Leonov},\ and\
  \citenamefont {Savrasov}}]{AFM82meV}%
  \BibitemOpen
  \bibfield  {author} {\bibinfo {author} {\bibfnamefont {X.}~\bibnamefont
  {Wan}}, \bibinfo {author} {\bibfnamefont {V.}~\bibnamefont {Ivanov}},
  \bibinfo {author} {\bibfnamefont {G.}~\bibnamefont {Resta}}, \bibinfo
  {author} {\bibfnamefont {I.}~\bibnamefont {Leonov}},\ and\ \bibinfo {author}
  {\bibfnamefont {S.~Y.}\ \bibnamefont {Savrasov}},\ }\bibfield  {title}
  {\bibinfo {title} {Exchange interactions and sensitivity of the {Ni} two-hole
  spin state to {Hund's} coupling in doped {NdNiO$_2$}},\ }\href
  {https://doi.org/10.1103/PhysRevB.103.075123} {\bibfield  {journal} {\bibinfo
   {journal} {Phys. Rev. B}\ }\textbf {\bibinfo {volume} {103}},\ \bibinfo
  {pages} {075123} (\bibinfo {year} {2021})}\BibitemShut {NoStop}%
\bibitem [{\citenamefont {Nomura}\ \emph {et~al.}(2020)\citenamefont {Nomura},
  \citenamefont {Nomoto}, \citenamefont {Hirayama},\ and\ \citenamefont
  {Arita}}]{AFM100meV}%
  \BibitemOpen
  \bibfield  {author} {\bibinfo {author} {\bibfnamefont {Y.}~\bibnamefont
  {Nomura}}, \bibinfo {author} {\bibfnamefont {T.}~\bibnamefont {Nomoto}},
  \bibinfo {author} {\bibfnamefont {M.}~\bibnamefont {Hirayama}},\ and\
  \bibinfo {author} {\bibfnamefont {R.}~\bibnamefont {Arita}},\ }\bibfield
  {title} {\bibinfo {title} {Magnetic exchange coupling in cuprate-analog
  ${d}^{9}$ nickelates},\ }\href
  {https://doi.org/10.1103/PhysRevResearch.2.043144} {\bibfield  {journal}
  {\bibinfo  {journal} {Phys. Rev. Research}\ }\textbf {\bibinfo {volume}
  {2}},\ \bibinfo {pages} {043144} (\bibinfo {year} {2020})}\BibitemShut
  {NoStop}%
\bibitem [{\citenamefont {Katukuri}\ \emph {et~al.}(2020)\citenamefont
  {Katukuri}, \citenamefont {Bogdanov}, \citenamefont {Weser}, \citenamefont
  {van~den Brink},\ and\ \citenamefont {Alavi}}]{AFM77meV}%
  \BibitemOpen
  \bibfield  {author} {\bibinfo {author} {\bibfnamefont {V.~M.}\ \bibnamefont
  {Katukuri}}, \bibinfo {author} {\bibfnamefont {N.~A.}\ \bibnamefont
  {Bogdanov}}, \bibinfo {author} {\bibfnamefont {O.}~\bibnamefont {Weser}},
  \bibinfo {author} {\bibfnamefont {J.}~\bibnamefont {van~den Brink}},\ and\
  \bibinfo {author} {\bibfnamefont {A.}~\bibnamefont {Alavi}},\ }\bibfield
  {title} {\bibinfo {title} {Electronic correlations and magnetic interactions
  in infinite-layer {${\mathrm{NdNiO}}_{2}$}},\ }\href
  {https://doi.org/10.1103/PhysRevB.102.241112} {\bibfield  {journal} {\bibinfo
   {journal} {Phys. Rev. B}\ }\textbf {\bibinfo {volume} {102}},\ \bibinfo
  {pages} {241112} (\bibinfo {year} {2020})}\BibitemShut {NoStop}%
\bibitem [{\citenamefont {Jung}\ \emph {et~al.}(2022)\citenamefont {Jung},
  \citenamefont {LaBollita}, \citenamefont {Pardo},\ and\ \citenamefont
  {Botana}}]{AFMBotana100}%
  \BibitemOpen
  \bibfield  {author} {\bibinfo {author} {\bibfnamefont {M.-C.}\ \bibnamefont
  {Jung}}, \bibinfo {author} {\bibfnamefont {H.}~\bibnamefont {LaBollita}},
  \bibinfo {author} {\bibfnamefont {V.}~\bibnamefont {Pardo}},\ and\ \bibinfo
  {author} {\bibfnamefont {A.~S.}\ \bibnamefont {Botana}},\ }\bibfield  {title}
  {\bibinfo {title} {Antiferromagnetic insulating state in layered nickelates
  at half filling},\ }\href {https://doi.org/10.1038/s41598-022-22176-2}
  {\bibfield  {journal} {\bibinfo  {journal} {Scientific Reports}\ }\textbf
  {\bibinfo {volume} {12}},\ \bibinfo {pages} {17864} (\bibinfo {year}
  {2022})}\BibitemShut {NoStop}%
\bibitem [{\citenamefont {Cui}\ \emph {et~al.}(2021)\citenamefont {Cui},
  \citenamefont {Li}, \citenamefont {Li}, \citenamefont {Zhu}, \citenamefont
  {Hu}, \citenamefont {Yang}, \citenamefont {Zhang}, \citenamefont {Yu},
  \citenamefont {Wen},\ and\ \citenamefont {Yu}}]{NMR}%
  \BibitemOpen
  \bibfield  {author} {\bibinfo {author} {\bibfnamefont {Y.}~\bibnamefont
  {Cui}}, \bibinfo {author} {\bibfnamefont {C.}~\bibnamefont {Li}}, \bibinfo
  {author} {\bibfnamefont {Q.}~\bibnamefont {Li}}, \bibinfo {author}
  {\bibfnamefont {X.}~\bibnamefont {Zhu}}, \bibinfo {author} {\bibfnamefont
  {Z.}~\bibnamefont {Hu}}, \bibinfo {author} {\bibfnamefont {Y.-F.}\
  \bibnamefont {Yang}}, \bibinfo {author} {\bibfnamefont {J.}~\bibnamefont
  {Zhang}}, \bibinfo {author} {\bibfnamefont {R.}~\bibnamefont {Yu}}, \bibinfo
  {author} {\bibfnamefont {H.-H.}\ \bibnamefont {Wen}},\ and\ \bibinfo {author}
  {\bibfnamefont {W.}~\bibnamefont {Yu}},\ }\bibfield  {title} {\bibinfo
  {title} {{NMR} evidence of antiferromagnetic spin fluctuations in
  {Nd$_{0.85}$Sr$_{0.15}$NiO$_2$}},\ }\href
  {https://doi.org/10.1088/0256-307x/38/6/067401} {\bibfield  {journal}
  {\bibinfo  {journal} {Chinese Physics Letters}\ }\textbf {\bibinfo {volume}
  {38}},\ \bibinfo {pages} {067401} (\bibinfo {year} {2021})}\BibitemShut
  {NoStop}%
\bibitem [{\citenamefont {Fu}\ \emph {et~al.}(2019)\citenamefont {Fu},
  \citenamefont {Wang}, \citenamefont {Cheng}, \citenamefont {Pei},
  \citenamefont {Zhou}, \citenamefont {Chen}, \citenamefont {Wang},
  \citenamefont {Zhao}, \citenamefont {Jiang}, \citenamefont {Liu},
  \citenamefont {Huang}, \citenamefont {Wang}, \citenamefont {Zhao},
  \citenamefont {Yu}, \citenamefont {Ye}, \citenamefont {Wang},\ and\
  \citenamefont {Mei}}]{raman}%
  \BibitemOpen
  \bibfield  {author} {\bibinfo {author} {\bibfnamefont {Y.}~\bibnamefont
  {Fu}}, \bibinfo {author} {\bibfnamefont {L.}~\bibnamefont {Wang}}, \bibinfo
  {author} {\bibfnamefont {H.}~\bibnamefont {Cheng}}, \bibinfo {author}
  {\bibfnamefont {S.}~\bibnamefont {Pei}}, \bibinfo {author} {\bibfnamefont
  {X.}~\bibnamefont {Zhou}}, \bibinfo {author} {\bibfnamefont {J.}~\bibnamefont
  {Chen}}, \bibinfo {author} {\bibfnamefont {S.}~\bibnamefont {Wang}}, \bibinfo
  {author} {\bibfnamefont {R.}~\bibnamefont {Zhao}}, \bibinfo {author}
  {\bibfnamefont {W.}~\bibnamefont {Jiang}}, \bibinfo {author} {\bibfnamefont
  {C.}~\bibnamefont {Liu}}, \bibinfo {author} {\bibfnamefont {M.}~\bibnamefont
  {Huang}}, \bibinfo {author} {\bibfnamefont {X.}~\bibnamefont {Wang}},
  \bibinfo {author} {\bibfnamefont {Y.}~\bibnamefont {Zhao}}, \bibinfo {author}
  {\bibfnamefont {D.}~\bibnamefont {Yu}}, \bibinfo {author} {\bibfnamefont
  {F.}~\bibnamefont {Ye}}, \bibinfo {author} {\bibfnamefont {S.}~\bibnamefont
  {Wang}},\ and\ \bibinfo {author} {\bibfnamefont {J.-W.}\ \bibnamefont
  {Mei}},\ }\href {https://doi.org/10.48550/ARXIV.1911.03177} {\bibinfo {title}
  {Core-level x-ray photoemission and raman spectroscopy studies on electronic
  structures in {Mott-Hubbard} type nickelate oxide {NdNiO$_2$}}} (\bibinfo
  {year} {2019}),\ \Eprint {https://arxiv.org/abs/1911.03177}
  {arXiv:1911.03177} \BibitemShut {NoStop}%
\bibitem [{\citenamefont {Lu}\ \emph {et~al.}(2021)\citenamefont {Lu},
  \citenamefont {Rossi}, \citenamefont {Nag}, \citenamefont {Osada},
  \citenamefont {Li}, \citenamefont {Lee}, \citenamefont {Wang}, \citenamefont
  {Garcia-Fernandez}, \citenamefont {Agrestini}, \citenamefont {Shen},
  \citenamefont {Been}, \citenamefont {Moritz}, \citenamefont {Devereaux},
  \citenamefont {Zaanen}, \citenamefont {Hwang}, \citenamefont {Zhou},\ and\
  \citenamefont {Lee}}]{RIXS}%
  \BibitemOpen
  \bibfield  {author} {\bibinfo {author} {\bibfnamefont {H.}~\bibnamefont
  {Lu}}, \bibinfo {author} {\bibfnamefont {M.}~\bibnamefont {Rossi}}, \bibinfo
  {author} {\bibfnamefont {A.}~\bibnamefont {Nag}}, \bibinfo {author}
  {\bibfnamefont {M.}~\bibnamefont {Osada}}, \bibinfo {author} {\bibfnamefont
  {D.~F.}\ \bibnamefont {Li}}, \bibinfo {author} {\bibfnamefont
  {K.}~\bibnamefont {Lee}}, \bibinfo {author} {\bibfnamefont {B.~Y.}\
  \bibnamefont {Wang}}, \bibinfo {author} {\bibfnamefont {M.}~\bibnamefont
  {Garcia-Fernandez}}, \bibinfo {author} {\bibfnamefont {S.}~\bibnamefont
  {Agrestini}}, \bibinfo {author} {\bibfnamefont {Z.~X.}\ \bibnamefont {Shen}},
  \bibinfo {author} {\bibfnamefont {E.~M.}\ \bibnamefont {Been}}, \bibinfo
  {author} {\bibfnamefont {B.}~\bibnamefont {Moritz}}, \bibinfo {author}
  {\bibfnamefont {T.~P.}\ \bibnamefont {Devereaux}}, \bibinfo {author}
  {\bibfnamefont {J.}~\bibnamefont {Zaanen}}, \bibinfo {author} {\bibfnamefont
  {H.~Y.}\ \bibnamefont {Hwang}}, \bibinfo {author} {\bibfnamefont {K.-J.}\
  \bibnamefont {Zhou}},\ and\ \bibinfo {author} {\bibfnamefont {W.~S.}\
  \bibnamefont {Lee}},\ }\bibfield  {title} {\bibinfo {title} {Magnetic
  excitations in infinite-layer nickelates},\ }\href
  {https://doi.org/10.1126/science.abd7726} {\bibfield  {journal} {\bibinfo
  {journal} {Science}\ }\textbf {\bibinfo {volume} {373}},\ \bibinfo {pages}
  {213} (\bibinfo {year} {2021})},\ \Eprint
  {https://arxiv.org/abs/https://science.sciencemag.org/content/373/6551/213.full.pdf}
  {https://science.sciencemag.org/content/373/6551/213.full.pdf} \BibitemShut
  {NoStop}%
\bibitem [{\citenamefont {Fowlie}\ \emph {et~al.}(2022)\citenamefont {Fowlie},
  \citenamefont {Hadjimichael}, \citenamefont {Martins}, \citenamefont {Li},
  \citenamefont {Osada}, \citenamefont {Wang}, \citenamefont {Lee},
  \citenamefont {Lee}, \citenamefont {Salman}, \citenamefont {Prokscha},
  \citenamefont {Triscone}, \citenamefont {Hwang},\ and\ \citenamefont
  {Suter}}]{muSR}%
  \BibitemOpen
  \bibfield  {author} {\bibinfo {author} {\bibfnamefont {J.}~\bibnamefont
  {Fowlie}}, \bibinfo {author} {\bibfnamefont {M.}~\bibnamefont
  {Hadjimichael}}, \bibinfo {author} {\bibfnamefont {M.~M.}\ \bibnamefont
  {Martins}}, \bibinfo {author} {\bibfnamefont {D.}~\bibnamefont {Li}},
  \bibinfo {author} {\bibfnamefont {M.}~\bibnamefont {Osada}}, \bibinfo
  {author} {\bibfnamefont {B.~Y.}\ \bibnamefont {Wang}}, \bibinfo {author}
  {\bibfnamefont {K.}~\bibnamefont {Lee}}, \bibinfo {author} {\bibfnamefont
  {Y.}~\bibnamefont {Lee}}, \bibinfo {author} {\bibfnamefont {Z.}~\bibnamefont
  {Salman}}, \bibinfo {author} {\bibfnamefont {T.}~\bibnamefont {Prokscha}},
  \bibinfo {author} {\bibfnamefont {J.-M.}\ \bibnamefont {Triscone}}, \bibinfo
  {author} {\bibfnamefont {H.~Y.}\ \bibnamefont {Hwang}},\ and\ \bibinfo
  {author} {\bibfnamefont {A.}~\bibnamefont {Suter}},\ }\bibfield  {title}
  {\bibinfo {title} {Intrinsic magnetism in superconducting infinite-layer
  nickelates},\ }\href {https://doi.org/10.1038/s41567-022-01684-y} {\bibfield
  {journal} {\bibinfo  {journal} {Nature Physics}\ }\textbf {\bibinfo {volume}
  {18}},\ \bibinfo {pages} {1043} (\bibinfo {year} {2022})}\BibitemShut
  {NoStop}%
\bibitem [{\citenamefont {Wu}\ \emph {et~al.}(2020{\natexlab{a}})\citenamefont
  {Wu}, \citenamefont {Di~Sante}, \citenamefont {Schwemmer}, \citenamefont
  {Hanke}, \citenamefont {Hwang}, \citenamefont {Raghu},\ and\ \citenamefont
  {Thomale}}]{DWavePredict}%
  \BibitemOpen
  \bibfield  {author} {\bibinfo {author} {\bibfnamefont {X.}~\bibnamefont
  {Wu}}, \bibinfo {author} {\bibfnamefont {D.}~\bibnamefont {Di~Sante}},
  \bibinfo {author} {\bibfnamefont {T.}~\bibnamefont {Schwemmer}}, \bibinfo
  {author} {\bibfnamefont {W.}~\bibnamefont {Hanke}}, \bibinfo {author}
  {\bibfnamefont {H.~Y.}\ \bibnamefont {Hwang}}, \bibinfo {author}
  {\bibfnamefont {S.}~\bibnamefont {Raghu}},\ and\ \bibinfo {author}
  {\bibfnamefont {R.}~\bibnamefont {Thomale}},\ }\bibfield  {title} {\bibinfo
  {title} {Robust ${d}_{{x}^{2}\ensuremath{-}{y}^{2}}$-wave superconductivity
  of infinite-layer nickelates},\ }\href
  {https://doi.org/10.1103/PhysRevB.101.060504} {\bibfield  {journal} {\bibinfo
   {journal} {Phys. Rev. B}\ }\textbf {\bibinfo {volume} {101}},\ \bibinfo
  {pages} {060504} (\bibinfo {year} {2020}{\natexlab{a}})}\BibitemShut
  {NoStop}%
\bibitem [{\citenamefont {Kitatani}\ \emph {et~al.}(2020)\citenamefont
  {Kitatani}, \citenamefont {Si}, \citenamefont {Janson}, \citenamefont
  {Arita}, \citenamefont {Zhong},\ and\ \citenamefont {Held}}]{DWaveJpn}%
  \BibitemOpen
  \bibfield  {author} {\bibinfo {author} {\bibfnamefont {M.}~\bibnamefont
  {Kitatani}}, \bibinfo {author} {\bibfnamefont {L.}~\bibnamefont {Si}},
  \bibinfo {author} {\bibfnamefont {O.}~\bibnamefont {Janson}}, \bibinfo
  {author} {\bibfnamefont {R.}~\bibnamefont {Arita}}, \bibinfo {author}
  {\bibfnamefont {Z.}~\bibnamefont {Zhong}},\ and\ \bibinfo {author}
  {\bibfnamefont {K.}~\bibnamefont {Held}},\ }\bibfield  {title} {\bibinfo
  {title} {Nickelate superconductors{\textemdash}a renaissance of the one-band
  {Hubbard} model},\ }\href {https://doi.org/10.1038/s41535-020-00260-y}
  {\bibfield  {journal} {\bibinfo  {journal} {npj Quantum Materials}\ }\textbf
  {\bibinfo {volume} {5}},\ \bibinfo {pages} {59} (\bibinfo {year}
  {2020})}\BibitemShut {NoStop}%
\bibitem [{\citenamefont {Chen}\ \emph
  {et~al.}(2022{\natexlab{c}})\citenamefont {Chen}, \citenamefont {Ma},
  \citenamefont {Sui}, \citenamefont {Liang}, \citenamefont {Huang},\ and\
  \citenamefont {Ma}}]{DWavePredict2}%
  \BibitemOpen
  \bibfield  {author} {\bibinfo {author} {\bibfnamefont {C.}~\bibnamefont
  {Chen}}, \bibinfo {author} {\bibfnamefont {R.}~\bibnamefont {Ma}}, \bibinfo
  {author} {\bibfnamefont {X.}~\bibnamefont {Sui}}, \bibinfo {author}
  {\bibfnamefont {Y.}~\bibnamefont {Liang}}, \bibinfo {author} {\bibfnamefont
  {B.}~\bibnamefont {Huang}},\ and\ \bibinfo {author} {\bibfnamefont
  {T.}~\bibnamefont {Ma}},\ }\bibfield  {title} {\bibinfo {title}
  {Antiferromagnetic fluctuations and dominant ${d}_{xy}$-wave pairing symmetry
  in nickelate-based superconductors},\ }\href
  {https://doi.org/10.1103/PhysRevB.106.195112} {\bibfield  {journal} {\bibinfo
   {journal} {Phys. Rev. B}\ }\textbf {\bibinfo {volume} {106}},\ \bibinfo
  {pages} {195112} (\bibinfo {year} {2022}{\natexlab{c}})}\BibitemShut
  {NoStop}%
\bibitem [{\citenamefont {Sakakibara}\ \emph {et~al.}(2020)\citenamefont
  {Sakakibara}, \citenamefont {Usui}, \citenamefont {Suzuki}, \citenamefont
  {Kotani}, \citenamefont {Aoki},\ and\ \citenamefont {Kuroki}}]{PossibleD}%
  \BibitemOpen
  \bibfield  {author} {\bibinfo {author} {\bibfnamefont {H.}~\bibnamefont
  {Sakakibara}}, \bibinfo {author} {\bibfnamefont {H.}~\bibnamefont {Usui}},
  \bibinfo {author} {\bibfnamefont {K.}~\bibnamefont {Suzuki}}, \bibinfo
  {author} {\bibfnamefont {T.}~\bibnamefont {Kotani}}, \bibinfo {author}
  {\bibfnamefont {H.}~\bibnamefont {Aoki}},\ and\ \bibinfo {author}
  {\bibfnamefont {K.}~\bibnamefont {Kuroki}},\ }\bibfield  {title} {\bibinfo
  {title} {Model construction and a possibility of cupratelike pairing in a new
  ${d}^{9}$ nickelate superconductor
  {$(\mathrm{Nd},\mathrm{Sr}){\mathrm{NiO}}_{2}$}},\ }\href
  {https://doi.org/10.1103/PhysRevLett.125.077003} {\bibfield  {journal}
  {\bibinfo  {journal} {Phys. Rev. Lett.}\ }\textbf {\bibinfo {volume} {125}},\
  \bibinfo {pages} {077003} (\bibinfo {year} {2020})}\BibitemShut {NoStop}%
\bibitem [{\citenamefont {Wang}\ \emph {et~al.}(2020)\citenamefont {Wang},
  \citenamefont {Zhang}, \citenamefont {Yang},\ and\ \citenamefont
  {Zhang}}]{SWaveDoping}%
  \BibitemOpen
  \bibfield  {author} {\bibinfo {author} {\bibfnamefont {Z.}~\bibnamefont
  {Wang}}, \bibinfo {author} {\bibfnamefont {G.-M.}\ \bibnamefont {Zhang}},
  \bibinfo {author} {\bibfnamefont {Y.-f.}\ \bibnamefont {Yang}},\ and\
  \bibinfo {author} {\bibfnamefont {F.-C.}\ \bibnamefont {Zhang}},\ }\bibfield
  {title} {\bibinfo {title} {Distinct pairing symmetries of superconductivity
  in infinite-layer nickelates},\ }\href
  {https://doi.org/10.1103/PhysRevB.102.220501} {\bibfield  {journal} {\bibinfo
   {journal} {Phys. Rev. B}\ }\textbf {\bibinfo {volume} {102}},\ \bibinfo
  {pages} {220501} (\bibinfo {year} {2020})}\BibitemShut {NoStop}%
\bibitem [{\citenamefont {Wu}\ \emph {et~al.}(2020{\natexlab{b}})\citenamefont
  {Wu}, \citenamefont {Jiang}, \citenamefont {Di~Sante}, \citenamefont {Hanke},
  \citenamefont {Schnyder}, \citenamefont {Hu},\ and\ \citenamefont
  {Thomale}}]{SWaveSurface}%
  \BibitemOpen
  \bibfield  {author} {\bibinfo {author} {\bibfnamefont {X.}~\bibnamefont
  {Wu}}, \bibinfo {author} {\bibfnamefont {K.}~\bibnamefont {Jiang}}, \bibinfo
  {author} {\bibfnamefont {D.}~\bibnamefont {Di~Sante}}, \bibinfo {author}
  {\bibfnamefont {W.}~\bibnamefont {Hanke}}, \bibinfo {author} {\bibfnamefont
  {A.~P.}\ \bibnamefont {Schnyder}}, \bibinfo {author} {\bibfnamefont
  {J.}~\bibnamefont {Hu}},\ and\ \bibinfo {author} {\bibfnamefont
  {R.}~\bibnamefont {Thomale}},\ }\href
  {https://doi.org/10.48550/ARXIV.2008.06009} {\bibinfo {title} {Surface
  $s$-wave superconductivity for oxide-terminated infinite-layer nickelates}}
  (\bibinfo {year} {2020}{\natexlab{b}}),\ \Eprint
  {https://arxiv.org/abs/2008.06009} {arXiv:2008.06009} \BibitemShut {NoStop}%
\bibitem [{\citenamefont {Chow}\ \emph {et~al.}(2022)\citenamefont {Chow},
  \citenamefont {Sudheesh}, \citenamefont {Nandi}, \citenamefont {Zeng},
  \citenamefont {Zhang}, \citenamefont {Du}, \citenamefont {Lim}, \citenamefont
  {Chia},\ and\ \citenamefont {Ariando}}]{sing_SD}%
  \BibitemOpen
  \bibfield  {author} {\bibinfo {author} {\bibfnamefont {L.~E.}\ \bibnamefont
  {Chow}}, \bibinfo {author} {\bibfnamefont {S.~K.}\ \bibnamefont {Sudheesh}},
  \bibinfo {author} {\bibfnamefont {P.}~\bibnamefont {Nandi}}, \bibinfo
  {author} {\bibfnamefont {S.~W.}\ \bibnamefont {Zeng}}, \bibinfo {author}
  {\bibfnamefont {Z.~T.}\ \bibnamefont {Zhang}}, \bibinfo {author}
  {\bibfnamefont {X.~M.}\ \bibnamefont {Du}}, \bibinfo {author} {\bibfnamefont
  {Z.~S.}\ \bibnamefont {Lim}}, \bibinfo {author} {\bibfnamefont {E.~E.~M.}\
  \bibnamefont {Chia}},\ and\ \bibinfo {author} {\bibfnamefont
  {A.}~\bibnamefont {Ariando}},\ }\href@noop {} {\bibinfo {title} {Pairing
  symmetry in infinite-layer nickelate superconductor}} (\bibinfo {year}
  {2022}),\ \Eprint {https://arxiv.org/abs/2201.10038} {arXiv:2201.10038}
  \BibitemShut {NoStop}%
\bibitem [{\citenamefont {Gu}\ \emph {et~al.}(2020)\citenamefont {Gu},
  \citenamefont {Li}, \citenamefont {Wan}, \citenamefont {Li}, \citenamefont
  {Guo}, \citenamefont {Yang}, \citenamefont {Li}, \citenamefont {Zhu},
  \citenamefont {Pan}, \citenamefont {Nie},\ and\ \citenamefont
  {Wen}}]{2020Tunnel}%
  \BibitemOpen
  \bibfield  {author} {\bibinfo {author} {\bibfnamefont {Q.}~\bibnamefont
  {Gu}}, \bibinfo {author} {\bibfnamefont {Y.}~\bibnamefont {Li}}, \bibinfo
  {author} {\bibfnamefont {S.}~\bibnamefont {Wan}}, \bibinfo {author}
  {\bibfnamefont {H.}~\bibnamefont {Li}}, \bibinfo {author} {\bibfnamefont
  {W.}~\bibnamefont {Guo}}, \bibinfo {author} {\bibfnamefont {H.}~\bibnamefont
  {Yang}}, \bibinfo {author} {\bibfnamefont {Q.}~\bibnamefont {Li}}, \bibinfo
  {author} {\bibfnamefont {X.}~\bibnamefont {Zhu}}, \bibinfo {author}
  {\bibfnamefont {X.}~\bibnamefont {Pan}}, \bibinfo {author} {\bibfnamefont
  {Y.}~\bibnamefont {Nie}},\ and\ \bibinfo {author} {\bibfnamefont {H.-H.}\
  \bibnamefont {Wen}},\ }\bibfield  {title} {\bibinfo {title} {Single particle
  tunneling spectrum of superconducting {Nd$_{1-x}$Sr$_x$NiO$_2$} thin films},\
  }\href {https://doi.org/10.1038/s41467-020-19908-1} {\bibfield  {journal}
  {\bibinfo  {journal} {Nature Communications}\ }\textbf {\bibinfo {volume}
  {11}},\ \bibinfo {pages} {6027} (\bibinfo {year} {2020})}\BibitemShut
  {NoStop}%
\bibitem [{\citenamefont {Harvey}\ \emph {et~al.}(2022)\citenamefont {Harvey},
  \citenamefont {Wang}, \citenamefont {Fowlie}, \citenamefont {Osada},
  \citenamefont {Lee}, \citenamefont {Lee}, \citenamefont {Li},\ and\
  \citenamefont {Hwang}}]{Hwang_SD}%
  \BibitemOpen
  \bibfield  {author} {\bibinfo {author} {\bibfnamefont {S.~P.}\ \bibnamefont
  {Harvey}}, \bibinfo {author} {\bibfnamefont {B.~Y.}\ \bibnamefont {Wang}},
  \bibinfo {author} {\bibfnamefont {J.}~\bibnamefont {Fowlie}}, \bibinfo
  {author} {\bibfnamefont {M.}~\bibnamefont {Osada}}, \bibinfo {author}
  {\bibfnamefont {K.}~\bibnamefont {Lee}}, \bibinfo {author} {\bibfnamefont
  {Y.}~\bibnamefont {Lee}}, \bibinfo {author} {\bibfnamefont {D.}~\bibnamefont
  {Li}},\ and\ \bibinfo {author} {\bibfnamefont {H.~Y.}\ \bibnamefont
  {Hwang}},\ }\href@noop {} {\bibinfo {title} {Evidence for nodal
  superconductivity in infinite-layer nickelates}} (\bibinfo {year} {2022}),\
  \Eprint {https://arxiv.org/abs/2201.12971} {arXiv:2201.12971} \BibitemShut
  {NoStop}%
\bibitem [{\citenamefont {Tsuei}\ \emph {et~al.}(1994)\citenamefont {Tsuei},
  \citenamefont {Kirtley}, \citenamefont {Chi}, \citenamefont {Yu-Jahnes},
  \citenamefont {Gupta}, \citenamefont {Shaw}, \citenamefont {Sun},\ and\
  \citenamefont {Ketchen}}]{triCrystal}%
  \BibitemOpen
  \bibfield  {author} {\bibinfo {author} {\bibfnamefont {C.~C.}\ \bibnamefont
  {Tsuei}}, \bibinfo {author} {\bibfnamefont {J.~R.}\ \bibnamefont {Kirtley}},
  \bibinfo {author} {\bibfnamefont {C.~C.}\ \bibnamefont {Chi}}, \bibinfo
  {author} {\bibfnamefont {L.~S.}\ \bibnamefont {Yu-Jahnes}}, \bibinfo {author}
  {\bibfnamefont {A.}~\bibnamefont {Gupta}}, \bibinfo {author} {\bibfnamefont
  {T.}~\bibnamefont {Shaw}}, \bibinfo {author} {\bibfnamefont {J.~Z.}\
  \bibnamefont {Sun}},\ and\ \bibinfo {author} {\bibfnamefont {M.~B.}\
  \bibnamefont {Ketchen}},\ }\bibfield  {title} {\bibinfo {title} {Pairing
  symmetry and flux quantization in a tricrystal superconducting ring of
  {YBa$_2$Cu$_3$O$_{7 -\delta}$}},\ }\href
  {https://doi.org/10.1103/PhysRevLett.73.593} {\bibfield  {journal} {\bibinfo
  {journal} {Phys. Rev. Lett.}\ }\textbf {\bibinfo {volume} {73}},\ \bibinfo
  {pages} {593} (\bibinfo {year} {1994})}\BibitemShut {NoStop}%
\bibitem [{\citenamefont {Kirtley}\ \emph {et~al.}(2016)\citenamefont
  {Kirtley}, \citenamefont {Paulius}, \citenamefont {Rosenberg}, \citenamefont
  {Palmstrom}, \citenamefont {Holland}, \citenamefont {Spanton}, \citenamefont
  {Schiessl}, \citenamefont {Jermain}, \citenamefont {Gibbons}, \citenamefont
  {Fung}, \citenamefont {Huber}, \citenamefont {Ralph}, \citenamefont
  {Ketchen}, \citenamefont {Gibson},\ and\ \citenamefont {Moler}}]{John2016}%
  \BibitemOpen
  \bibfield  {author} {\bibinfo {author} {\bibfnamefont {J.~R.}\ \bibnamefont
  {Kirtley}}, \bibinfo {author} {\bibfnamefont {L.}~\bibnamefont {Paulius}},
  \bibinfo {author} {\bibfnamefont {A.~J.}\ \bibnamefont {Rosenberg}}, \bibinfo
  {author} {\bibfnamefont {J.~C.}\ \bibnamefont {Palmstrom}}, \bibinfo {author}
  {\bibfnamefont {C.~M.}\ \bibnamefont {Holland}}, \bibinfo {author}
  {\bibfnamefont {E.~M.}\ \bibnamefont {Spanton}}, \bibinfo {author}
  {\bibfnamefont {D.}~\bibnamefont {Schiessl}}, \bibinfo {author}
  {\bibfnamefont {C.~L.}\ \bibnamefont {Jermain}}, \bibinfo {author}
  {\bibfnamefont {J.}~\bibnamefont {Gibbons}}, \bibinfo {author} {\bibfnamefont
  {Y.-K.-K.}\ \bibnamefont {Fung}}, \bibinfo {author} {\bibfnamefont {M.~E.}\
  \bibnamefont {Huber}}, \bibinfo {author} {\bibfnamefont {D.~C.}\ \bibnamefont
  {Ralph}}, \bibinfo {author} {\bibfnamefont {M.~B.}\ \bibnamefont {Ketchen}},
  \bibinfo {author} {\bibfnamefont {G.~W.}\ \bibnamefont {Gibson}},\ and\
  \bibinfo {author} {\bibfnamefont {K.~A.}\ \bibnamefont {Moler}},\ }\bibfield
  {title} {\bibinfo {title} {Scanning {SQUID} susceptometers with sub-micron
  spatial resolution},\ }\href {https://doi.org/10.1063/1.4961982} {\bibfield
  {journal} {\bibinfo  {journal} {Review of Scientific Instruments}\ }\textbf
  {\bibinfo {volume} {87}},\ \bibinfo {pages} {093702} (\bibinfo {year}
  {2016})},\ \Eprint {https://arxiv.org/abs/https://doi.org/10.1063/1.4961982}
  {https://doi.org/10.1063/1.4961982} \BibitemShut {NoStop}%
\bibitem [{\citenamefont {Bert}\ \emph {et~al.}(2011)\citenamefont {Bert},
  \citenamefont {Kalisky}, \citenamefont {Bell}, \citenamefont {Kim},
  \citenamefont {Hikita}, \citenamefont {Hwang},\ and\ \citenamefont
  {Moler}}]{LAOSTO}%
  \BibitemOpen
  \bibfield  {author} {\bibinfo {author} {\bibfnamefont {J.~A.}\ \bibnamefont
  {Bert}}, \bibinfo {author} {\bibfnamefont {B.}~\bibnamefont {Kalisky}},
  \bibinfo {author} {\bibfnamefont {C.}~\bibnamefont {Bell}}, \bibinfo {author}
  {\bibfnamefont {M.}~\bibnamefont {Kim}}, \bibinfo {author} {\bibfnamefont
  {Y.}~\bibnamefont {Hikita}}, \bibinfo {author} {\bibfnamefont {H.~Y.}\
  \bibnamefont {Hwang}},\ and\ \bibinfo {author} {\bibfnamefont {K.~A.}\
  \bibnamefont {Moler}},\ }\bibfield  {title} {\bibinfo {title} {Direct imaging
  of the coexistence of ferromagnetism and superconductivity
  at the {LaAlO$_3$/SrTiO$_3$} interface},\ }\href
  {https://doi.org/10.1038/nphys2079} {\bibfield  {journal} {\bibinfo
  {journal} {Nature Physics}\ }\textbf {\bibinfo {volume} {7}},\ \bibinfo
  {pages} {767} (\bibinfo {year} {2011})}\BibitemShut {NoStop}%
\bibitem [{\citenamefont {Tafuri}\ \emph {et~al.}(2004)\citenamefont {Tafuri},
  \citenamefont {Kirtley}, \citenamefont {Medaglia}, \citenamefont {Orgiani},\
  and\ \citenamefont {Balestrino}}]{PearlVortex}%
  \BibitemOpen
  \bibfield  {author} {\bibinfo {author} {\bibfnamefont {F.}~\bibnamefont
  {Tafuri}}, \bibinfo {author} {\bibfnamefont {J.~R.}\ \bibnamefont {Kirtley}},
  \bibinfo {author} {\bibfnamefont {P.~G.}\ \bibnamefont {Medaglia}}, \bibinfo
  {author} {\bibfnamefont {P.}~\bibnamefont {Orgiani}},\ and\ \bibinfo {author}
  {\bibfnamefont {G.}~\bibnamefont {Balestrino}},\ }\bibfield  {title}
  {\bibinfo {title} {Magnetic imaging of pearl vortices in artificially layered
  {$({\mathrm{Ba}}_{0.9}{\mathrm{Nd}}_{0.1}{\mathrm{CuO}}_{2+x}{)}_{m}/({\mathrm{CaCuO}}_{2}{)}_{n}$}
  systems},\ }\href {https://doi.org/10.1103/PhysRevLett.92.157006} {\bibfield
  {journal} {\bibinfo  {journal} {Phys. Rev. Lett.}\ }\textbf {\bibinfo
  {volume} {92}},\ \bibinfo {pages} {157006} (\bibinfo {year}
  {2004})}\BibitemShut {NoStop}%
\bibitem [{\citenamefont {Davis}\ \emph {et~al.}(2018)\citenamefont {Davis},
  \citenamefont {Ullah}, \citenamefont {Adamo}, \citenamefont {Watson},
  \citenamefont {Kirtley}, \citenamefont {Beasley}, \citenamefont {Kivelson},\
  and\ \citenamefont {Moler}}]{Sam}%
  \BibitemOpen
  \bibfield  {author} {\bibinfo {author} {\bibfnamefont {S.~I.}\ \bibnamefont
  {Davis}}, \bibinfo {author} {\bibfnamefont {R.~R.}\ \bibnamefont {Ullah}},
  \bibinfo {author} {\bibfnamefont {C.}~\bibnamefont {Adamo}}, \bibinfo
  {author} {\bibfnamefont {C.~A.}\ \bibnamefont {Watson}}, \bibinfo {author}
  {\bibfnamefont {J.~R.}\ \bibnamefont {Kirtley}}, \bibinfo {author}
  {\bibfnamefont {M.~R.}\ \bibnamefont {Beasley}}, \bibinfo {author}
  {\bibfnamefont {S.~A.}\ \bibnamefont {Kivelson}},\ and\ \bibinfo {author}
  {\bibfnamefont {K.~A.}\ \bibnamefont {Moler}},\ }\bibfield  {title} {\bibinfo
  {title} {Spatially modulated susceptibility in thin film
  {${\mathrm{La}}_{2\ensuremath{-}x}{\mathrm{Ba}}_{x}{\mathrm{CuO}}_{4}$}},\
  }\href {https://doi.org/10.1103/PhysRevB.98.014506} {\bibfield  {journal}
  {\bibinfo  {journal} {Phys. Rev. B}\ }\textbf {\bibinfo {volume} {98}},\
  \bibinfo {pages} {014506} (\bibinfo {year} {2018})}\BibitemShut {NoStop}%
\bibitem [{\citenamefont {Lee}\ \emph {et~al.}(2020)\citenamefont {Lee},
  \citenamefont {Goodge}, \citenamefont {Li}, \citenamefont {Osada},
  \citenamefont {Wang}, \citenamefont {Cui}, \citenamefont {Kourkoutis},\ and\
  \citenamefont {Hwang}}]{2020Synthesis}%
  \BibitemOpen
  \bibfield  {author} {\bibinfo {author} {\bibfnamefont {K.}~\bibnamefont
  {Lee}}, \bibinfo {author} {\bibfnamefont {B.~H.}\ \bibnamefont {Goodge}},
  \bibinfo {author} {\bibfnamefont {D.}~\bibnamefont {Li}}, \bibinfo {author}
  {\bibfnamefont {M.}~\bibnamefont {Osada}}, \bibinfo {author} {\bibfnamefont
  {B.~Y.}\ \bibnamefont {Wang}}, \bibinfo {author} {\bibfnamefont
  {Y.}~\bibnamefont {Cui}}, \bibinfo {author} {\bibfnamefont {L.~F.}\
  \bibnamefont {Kourkoutis}},\ and\ \bibinfo {author} {\bibfnamefont {H.~Y.}\
  \bibnamefont {Hwang}},\ }\bibfield  {title} {\bibinfo {title} {Aspects of the
  synthesis of thin film superconducting infinite-layer nickelates},\ }\href
  {https://doi.org/10.1063/5.0005103} {\bibfield  {journal} {\bibinfo
  {journal} {{APL} Materials}\ }\textbf {\bibinfo {volume} {8}},\ \bibinfo
  {pages} {041107} (\bibinfo {year} {2020})}\BibitemShut {NoStop}%
\bibitem [{Sup()}]{Supplemental}%
  \BibitemOpen
  \href@noop {} {\bibinfo  {journal} {See Supplemental Material at [URL will be
  inserted by publisher] for sample characterization, optical images of line-
  and cross-shape inhomogeneities, characterization of magnetic inclusions,
  temperature-dependent superfluid density fitting, and Pearl vortex fitting}\
  }\BibitemShut {NoStop}%
\bibitem [{\citenamefont {Noad}\ \emph {et~al.}(2018)\citenamefont {Noad},
  \citenamefont {Watson}, \citenamefont {Inoue}, \citenamefont {Kim},
  \citenamefont {Sato}, \citenamefont {Bell}, \citenamefont {Hwang},
  \citenamefont {Kirtley},\ and\ \citenamefont {Moler}}]{defect}%
  \BibitemOpen
\bibfield  {journal} {  }\bibfield  {author} {\bibinfo {author} {\bibfnamefont
  {H.}~\bibnamefont {Noad}}, \bibinfo {author} {\bibfnamefont {C.~A.}\
  \bibnamefont {Watson}}, \bibinfo {author} {\bibfnamefont {H.}~\bibnamefont
  {Inoue}}, \bibinfo {author} {\bibfnamefont {M.}~\bibnamefont {Kim}}, \bibinfo
  {author} {\bibfnamefont {H.~K.}\ \bibnamefont {Sato}}, \bibinfo {author}
  {\bibfnamefont {C.}~\bibnamefont {Bell}}, \bibinfo {author} {\bibfnamefont
  {H.~Y.}\ \bibnamefont {Hwang}}, \bibinfo {author} {\bibfnamefont {J.~R.}\
  \bibnamefont {Kirtley}},\ and\ \bibinfo {author} {\bibfnamefont {K.~A.}\
  \bibnamefont {Moler}},\ }\bibfield  {title} {\bibinfo {title} {Observation of
  signatures of subresolution defects in two-dimensional superconductors with a
  scanning {SQUID}},\ }\href {https://doi.org/10.1103/PhysRevB.98.064510}
  {\bibfield  {journal} {\bibinfo  {journal} {Phys. Rev. B}\ }\textbf {\bibinfo
  {volume} {98}},\ \bibinfo {pages} {064510} (\bibinfo {year}
  {2018})}\BibitemShut {NoStop}%
\bibitem [{\citenamefont {Bi}\ \emph {et~al.}(2004)\citenamefont {Bi},
  \citenamefont {Li}, \citenamefont {Zhang},\ and\ \citenamefont
  {Du}}]{2004NiO}%
  \BibitemOpen
  \bibfield  {author} {\bibinfo {author} {\bibfnamefont {H.}~\bibnamefont
  {Bi}}, \bibinfo {author} {\bibfnamefont {S.}~\bibnamefont {Li}}, \bibinfo
  {author} {\bibfnamefont {Y.}~\bibnamefont {Zhang}},\ and\ \bibinfo {author}
  {\bibfnamefont {Y.}~\bibnamefont {Du}},\ }\bibfield  {title} {\bibinfo
  {title} {Ferromagnetic-like behavior of ultrafine {NiO} nanocrystallites},\
  }\href {https://doi.org/10.1016/j.jmmm.2003.11.017} {\bibfield  {journal}
  {\bibinfo  {journal} {Journal of Magnetism and Magnetic Materials}\ }\textbf
  {\bibinfo {volume} {277}},\ \bibinfo {pages} {363} (\bibinfo {year}
  {2004})}\BibitemShut {NoStop}%
\bibitem [{\citenamefont {Winkler}\ \emph {et~al.}(2005)\citenamefont
  {Winkler}, \citenamefont {Zysler}, \citenamefont {Mansilla},\ and\
  \citenamefont {Fiorani}}]{2005NiO}%
  \BibitemOpen
  \bibfield  {author} {\bibinfo {author} {\bibfnamefont {E.}~\bibnamefont
  {Winkler}}, \bibinfo {author} {\bibfnamefont {R.~D.}\ \bibnamefont {Zysler}},
  \bibinfo {author} {\bibfnamefont {M.~V.}\ \bibnamefont {Mansilla}},\ and\
  \bibinfo {author} {\bibfnamefont {D.}~\bibnamefont {Fiorani}},\ }\bibfield
  {title} {\bibinfo {title} {Surface anisotropy effects in {NiO}
  nanoparticles},\ }\href {https://doi.org/10.1103/PhysRevB.72.132409}
  {\bibfield  {journal} {\bibinfo  {journal} {Phys. Rev. B}\ }\textbf {\bibinfo
  {volume} {72}},\ \bibinfo {pages} {132409} (\bibinfo {year}
  {2005})}\BibitemShut {NoStop}%
\bibitem [{\citenamefont {Ghosh}\ \emph {et~al.}(2006)\citenamefont {Ghosh},
  \citenamefont {Biswas}, \citenamefont {Sundaresan},\ and\ \citenamefont
  {Rao}}]{2006NiO}%
  \BibitemOpen
  \bibfield  {author} {\bibinfo {author} {\bibfnamefont {M.}~\bibnamefont
  {Ghosh}}, \bibinfo {author} {\bibfnamefont {K.}~\bibnamefont {Biswas}},
  \bibinfo {author} {\bibfnamefont {A.}~\bibnamefont {Sundaresan}},\ and\
  \bibinfo {author} {\bibfnamefont {C.~N.~R.}\ \bibnamefont {Rao}},\ }\bibfield
   {title} {\bibinfo {title} {{MnO and NiO} nanoparticles: synthesis and
  magnetic properties},\ }\href {https://doi.org/10.1039/B511920K} {\bibfield
  {journal} {\bibinfo  {journal} {J. Mater. Chem.}\ }\textbf {\bibinfo {volume}
  {16}},\ \bibinfo {pages} {106} (\bibinfo {year} {2006})}\BibitemShut
  {NoStop}%
\bibitem [{\citenamefont {Sharma}\ \emph {et~al.}(2009)\citenamefont {Sharma},
  \citenamefont {Vargas}, \citenamefont {Biasi}, \citenamefont {B{\'{e}}ron},
  \citenamefont {Knobel}, \citenamefont {Pirota}, \citenamefont {Meneses},
  \citenamefont {Kumar}, \citenamefont {Lee}, \citenamefont {Pagliuso},\ and\
  \citenamefont {Rettori}}]{2009NiO}%
  \BibitemOpen
  \bibfield  {author} {\bibinfo {author} {\bibfnamefont {S.~K.}\ \bibnamefont
  {Sharma}}, \bibinfo {author} {\bibfnamefont {J.~M.}\ \bibnamefont {Vargas}},
  \bibinfo {author} {\bibfnamefont {E.~D.}\ \bibnamefont {Biasi}}, \bibinfo
  {author} {\bibfnamefont {F.}~\bibnamefont {B{\'{e}}ron}}, \bibinfo {author}
  {\bibfnamefont {M.}~\bibnamefont {Knobel}}, \bibinfo {author} {\bibfnamefont
  {K.~R.}\ \bibnamefont {Pirota}}, \bibinfo {author} {\bibfnamefont {C.~T.}\
  \bibnamefont {Meneses}}, \bibinfo {author} {\bibfnamefont {S.}~\bibnamefont
  {Kumar}}, \bibinfo {author} {\bibfnamefont {C.~G.}\ \bibnamefont {Lee}},
  \bibinfo {author} {\bibfnamefont {P.~G.}\ \bibnamefont {Pagliuso}},\ and\
  \bibinfo {author} {\bibfnamefont {C.}~\bibnamefont {Rettori}},\ }\bibfield
  {title} {\bibinfo {title} {The nature and enhancement of magnetic surface
  contribution in model {NiO} nanoparticles},\ }\href
  {https://doi.org/10.1088/0957-4484/21/3/035602} {\bibfield  {journal}
  {\bibinfo  {journal} {Nanotechnology}\ }\textbf {\bibinfo {volume} {21}},\
  \bibinfo {pages} {035602} (\bibinfo {year} {2009})}\BibitemShut {NoStop}%
\bibitem [{\citenamefont {Tadic}\ \emph {et~al.}(2015)\citenamefont {Tadic},
  \citenamefont {Nikolic}, \citenamefont {Panjan},\ and\ \citenamefont
  {Blake}}]{NiO2015}%
  \BibitemOpen
  \bibfield  {author} {\bibinfo {author} {\bibfnamefont {M.}~\bibnamefont
  {Tadic}}, \bibinfo {author} {\bibfnamefont {D.}~\bibnamefont {Nikolic}},
  \bibinfo {author} {\bibfnamefont {M.}~\bibnamefont {Panjan}},\ and\ \bibinfo
  {author} {\bibfnamefont {G.~R.}\ \bibnamefont {Blake}},\ }\bibfield  {title}
  {\bibinfo {title} {Magnetic properties of {NiO} (nickel oxide) nanoparticles:
  Blocking temperature and {Neel} temperature},\ }\href
  {https://doi.org/https://doi.org/10.1016/j.jallcom.2015.06.027} {\bibfield
  {journal} {\bibinfo  {journal} {Journal of Alloys and Compounds}\ }\textbf
  {\bibinfo {volume} {647}},\ \bibinfo {pages} {1061} (\bibinfo {year}
  {2015})}\BibitemShut {NoStop}%
\bibitem [{\citenamefont {Luis}\ \emph {et~al.}(1999)\citenamefont {Luis},
  \citenamefont {del Barco}, \citenamefont {Hern\'andez}, \citenamefont
  {Remiro}, \citenamefont {Bartolom\'e},\ and\ \citenamefont
  {Tejada}}]{superpara}%
  \BibitemOpen
  \bibfield  {author} {\bibinfo {author} {\bibfnamefont {F.}~\bibnamefont
  {Luis}}, \bibinfo {author} {\bibfnamefont {E.}~\bibnamefont {del Barco}},
  \bibinfo {author} {\bibfnamefont {J.~M.}\ \bibnamefont {Hern\'andez}},
  \bibinfo {author} {\bibfnamefont {E.}~\bibnamefont {Remiro}}, \bibinfo
  {author} {\bibfnamefont {J.}~\bibnamefont {Bartolom\'e}},\ and\ \bibinfo
  {author} {\bibfnamefont {J.}~\bibnamefont {Tejada}},\ }\bibfield  {title}
  {\bibinfo {title} {Resonant spin tunneling in small antiferromagnetic
  particles},\ }\href {https://doi.org/10.1103/PhysRevB.59.11837} {\bibfield
  {journal} {\bibinfo  {journal} {Phys. Rev. B}\ }\textbf {\bibinfo {volume}
  {59}},\ \bibinfo {pages} {11837} (\bibinfo {year} {1999})}\BibitemShut
  {NoStop}%
\bibitem [{\citenamefont {Kirtley}\ \emph {et~al.}(2012)\citenamefont
  {Kirtley}, \citenamefont {Kalisky}, \citenamefont {Bert}, \citenamefont
  {Bell}, \citenamefont {Kim}, \citenamefont {Hikita}, \citenamefont {Hwang},
  \citenamefont {Ngai}, \citenamefont {Segal}, \citenamefont {Walker},
  \citenamefont {Ahn},\ and\ \citenamefont {Moler}}]{JohnTD}%
  \BibitemOpen
  \bibfield  {author} {\bibinfo {author} {\bibfnamefont {J.~R.}\ \bibnamefont
  {Kirtley}}, \bibinfo {author} {\bibfnamefont {B.}~\bibnamefont {Kalisky}},
  \bibinfo {author} {\bibfnamefont {J.~A.}\ \bibnamefont {Bert}}, \bibinfo
  {author} {\bibfnamefont {C.}~\bibnamefont {Bell}}, \bibinfo {author}
  {\bibfnamefont {M.}~\bibnamefont {Kim}}, \bibinfo {author} {\bibfnamefont
  {Y.}~\bibnamefont {Hikita}}, \bibinfo {author} {\bibfnamefont {H.~Y.}\
  \bibnamefont {Hwang}}, \bibinfo {author} {\bibfnamefont {J.~H.}\ \bibnamefont
  {Ngai}}, \bibinfo {author} {\bibfnamefont {Y.}~\bibnamefont {Segal}},
  \bibinfo {author} {\bibfnamefont {F.~J.}\ \bibnamefont {Walker}}, \bibinfo
  {author} {\bibfnamefont {C.~H.}\ \bibnamefont {Ahn}},\ and\ \bibinfo {author}
  {\bibfnamefont {K.~A.}\ \bibnamefont {Moler}},\ }\bibfield  {title} {\bibinfo
  {title} {Scanning {SQUID} susceptometry of a paramagnetic superconductor},\
  }\href {https://doi.org/10.1103/PhysRevB.85.224518} {\bibfield  {journal}
  {\bibinfo  {journal} {Phys. Rev. B}\ }\textbf {\bibinfo {volume} {85}},\
  \bibinfo {pages} {224518} (\bibinfo {year} {2012})}\BibitemShut {NoStop}%
\bibitem [{\citenamefont {Ortiz}\ \emph {et~al.}(2022)\citenamefont {Ortiz},
  \citenamefont {Puphal}, \citenamefont {Klett}, \citenamefont {Hotz},
  \citenamefont {Kremer}, \citenamefont {Trepka}, \citenamefont {Hemmida},
  \citenamefont {von Nidda}, \citenamefont {Isobe}, \citenamefont {Khasanov},
  \citenamefont {Luetkens}, \citenamefont {Hansmann}, \citenamefont {Keimer},
  \citenamefont {Sch\"afer},\ and\ \citenamefont {Hepting}}]{powderSpinGlass}%
  \BibitemOpen
  \bibfield  {author} {\bibinfo {author} {\bibfnamefont {R.~A.}\ \bibnamefont
  {Ortiz}}, \bibinfo {author} {\bibfnamefont {P.}~\bibnamefont {Puphal}},
  \bibinfo {author} {\bibfnamefont {M.}~\bibnamefont {Klett}}, \bibinfo
  {author} {\bibfnamefont {F.}~\bibnamefont {Hotz}}, \bibinfo {author}
  {\bibfnamefont {R.~K.}\ \bibnamefont {Kremer}}, \bibinfo {author}
  {\bibfnamefont {H.}~\bibnamefont {Trepka}}, \bibinfo {author} {\bibfnamefont
  {M.}~\bibnamefont {Hemmida}}, \bibinfo {author} {\bibfnamefont {H.-A.~K.}\
  \bibnamefont {von Nidda}}, \bibinfo {author} {\bibfnamefont {M.}~\bibnamefont
  {Isobe}}, \bibinfo {author} {\bibfnamefont {R.}~\bibnamefont {Khasanov}},
  \bibinfo {author} {\bibfnamefont {H.}~\bibnamefont {Luetkens}}, \bibinfo
  {author} {\bibfnamefont {P.}~\bibnamefont {Hansmann}}, \bibinfo {author}
  {\bibfnamefont {B.}~\bibnamefont {Keimer}}, \bibinfo {author} {\bibfnamefont
  {T.}~\bibnamefont {Sch\"afer}},\ and\ \bibinfo {author} {\bibfnamefont
  {M.}~\bibnamefont {Hepting}},\ }\bibfield  {title} {\bibinfo {title}
  {Magnetic correlations in infinite-layer nickelates: An experimental and
  theoretical multimethod study},\ }\href
  {https://doi.org/10.1103/PhysRevResearch.4.023093} {\bibfield  {journal}
  {\bibinfo  {journal} {Phys. Rev. Res.}\ }\textbf {\bibinfo {volume} {4}},\
  \bibinfo {pages} {023093} (\bibinfo {year} {2022})}\BibitemShut {NoStop}%
\bibitem [{\citenamefont {Kamber}\ \emph {et~al.}(2020)\citenamefont {Kamber},
  \citenamefont {Bergman}, \citenamefont {Eich}, \citenamefont {Iu{\c{s}}an},
  \citenamefont {Steinbrecher}, \citenamefont {Hauptmann}, \citenamefont
  {Nordström}, \citenamefont {Katsnelson}, \citenamefont {Wegner},
  \citenamefont {Eriksson},\ and\ \citenamefont {Khajetoorians}}]{glass_STM}%
  \BibitemOpen
  \bibfield  {author} {\bibinfo {author} {\bibfnamefont {U.}~\bibnamefont
  {Kamber}}, \bibinfo {author} {\bibfnamefont {A.}~\bibnamefont {Bergman}},
  \bibinfo {author} {\bibfnamefont {A.}~\bibnamefont {Eich}}, \bibinfo {author}
  {\bibfnamefont {D.}~\bibnamefont {Iu{\c{s}}an}}, \bibinfo {author}
  {\bibfnamefont {M.}~\bibnamefont {Steinbrecher}}, \bibinfo {author}
  {\bibfnamefont {N.}~\bibnamefont {Hauptmann}}, \bibinfo {author}
  {\bibfnamefont {L.}~\bibnamefont {Nordström}}, \bibinfo {author}
  {\bibfnamefont {M.~I.}\ \bibnamefont {Katsnelson}}, \bibinfo {author}
  {\bibfnamefont {D.}~\bibnamefont {Wegner}}, \bibinfo {author} {\bibfnamefont
  {O.}~\bibnamefont {Eriksson}},\ and\ \bibinfo {author} {\bibfnamefont
  {A.~A.}\ \bibnamefont {Khajetoorians}},\ }\bibfield  {title} {\bibinfo
  {title} {Self-induced spin glass state in elemental and crystalline
  neodymium},\ }\href {https://doi.org/10.1126/science.aay6757} {\bibfield
  {journal} {\bibinfo  {journal} {Science}\ }\textbf {\bibinfo {volume}
  {368}},\ \bibinfo {pages} {eaay6757} (\bibinfo {year} {2020})}\BibitemShut
  {NoStop}%
\bibitem [{\citenamefont {Wu}\ \emph {et~al.}(2006)\citenamefont {Wu},
  \citenamefont {Israel}, \citenamefont {Hur}, \citenamefont {Park},
  \citenamefont {Cheong},\ and\ \citenamefont {de~Lozanne}}]{glass_AFM}%
  \BibitemOpen
  \bibfield  {author} {\bibinfo {author} {\bibfnamefont {W.}~\bibnamefont
  {Wu}}, \bibinfo {author} {\bibfnamefont {C.}~\bibnamefont {Israel}}, \bibinfo
  {author} {\bibfnamefont {N.}~\bibnamefont {Hur}}, \bibinfo {author}
  {\bibfnamefont {S.}~\bibnamefont {Park}}, \bibinfo {author} {\bibfnamefont
  {S.-W.}\ \bibnamefont {Cheong}},\ and\ \bibinfo {author} {\bibfnamefont
  {A.}~\bibnamefont {de~Lozanne}},\ }\bibfield  {title} {\bibinfo {title}
  {Magnetic imaging of a supercooling glass transition in a weakly disordered
  ferromagnet},\ }\href {https://doi.org/10.1038/nmat1743} {\bibfield
  {journal} {\bibinfo  {journal} {Nature Materials}\ }\textbf {\bibinfo
  {volume} {5}},\ \bibinfo {pages} {881} (\bibinfo {year} {2006})}\BibitemShut
  {NoStop}%
\bibitem [{\citenamefont {Prozorov}\ and\ \citenamefont
  {Giannetta}(2006)}]{SD_formula}%
  \BibitemOpen
  \bibfield  {author} {\bibinfo {author} {\bibfnamefont {R.}~\bibnamefont
  {Prozorov}}\ and\ \bibinfo {author} {\bibfnamefont {R.~W.}\ \bibnamefont
  {Giannetta}},\ }\bibfield  {title} {\bibinfo {title} {Magnetic penetration
  depth in unconventional superconductors},\ }\href
  {https://doi.org/10.1088/0953-2048/19/8/r01} {\bibfield  {journal} {\bibinfo
  {journal} {Superconductor Science and Technology}\ }\textbf {\bibinfo
  {volume} {19}},\ \bibinfo {pages} {R41} (\bibinfo {year} {2006})}\BibitemShut
  {NoStop}%
\bibitem [{\citenamefont {Hirschfeld}\ and\ \citenamefont
  {Goldenfeld}(1993)}]{CrossOver}%
  \BibitemOpen
  \bibfield  {author} {\bibinfo {author} {\bibfnamefont {P.~J.}\ \bibnamefont
  {Hirschfeld}}\ and\ \bibinfo {author} {\bibfnamefont {N.}~\bibnamefont
  {Goldenfeld}},\ }\bibfield  {title} {\bibinfo {title} {Effect of strong
  scattering on the low-temperature penetration depth of a d-wave
  superconductor},\ }\href {https://doi.org/10.1103/PhysRevB.48.4219}
  {\bibfield  {journal} {\bibinfo  {journal} {Phys. Rev. B}\ }\textbf {\bibinfo
  {volume} {48}},\ \bibinfo {pages} {4219} (\bibinfo {year}
  {1993})}\BibitemShut {NoStop}%
\end{thebibliography}%
\nocite{defect}

\end{document}